\documentclass[twocolumn,times]{aastex63}

\received{XXXX XX, 2020}

\shorttitle{The supermassive black hole masses}
\shortauthors{Yu et al.}

\usepackage{graphicx}
\usepackage{multirow}
\usepackage{color}
\usepackage[figuresright]{rotating}
\usepackage{longtable}
\usepackage{mathrsfs,amsmath}
\usepackage{epstopdf}
\newsavebox{\tablebox}
\usepackage{hyperref}
\usepackage{times}
\usepackage{longtable}
\usepackage{array}
\usepackage{booktabs}

\newcommand{\blue}[1]{\textcolor{blue}{#1}}

\newcommand{\lv}{\ifmmode L_{5100} \else $L_{5100}$\ \fi}
\newcommand{\kms}{\ifmmode {\rm km\ s}^{-1} \else km s$^{-1}$\ \fi}
\newcommand{\ergs}{\ifmmode {\rm erg\ s}^{-1} \else erg s$^{-1}$\ \fi}
\newcommand{\lb}{\ifmmode L_{\rm Bol} \else $L_{\rm Bol}$\ \fi}
\newcommand{\ledd}{\ifmmode L_{\rm Edd} \else $L_{\rm Edd}$\ \fi}
\newcommand{\hb}{\ifmmode H\beta \else H$\beta$\ \fi}
\newcommand{\ha}{\ifmmode H\alpha \else H$\alpha$\ \fi}

\newcommand{\civ}{C {\sc iv}}

\newcommand{\mbh}{\ifmmode M_{\rm BH}  \else $M_{\rm BH}$\ \fi}
\newcommand{\msun}{M_{\odot}}
\newcommand{\rfe}{\ifmmode R_{\rm Fe} \else $R_{\rm Fe}$\ \fi}
\newcommand{\sst}{\ifmmode \sigma_{\rm \ast}\else $\sigma_{\rm \ast}$\ \fi}
\newcommand{\dhb}{\ifmmode D_{\rm H\beta} \else $D_{\rm H\beta}$\ \fi}
\newcommand{\leddR}{\ifmmode L_{\rm Bol}/L_{\rm Edd} \else $L_{\rm Bol}/L_{\rm Edd}$\ \fi}
\newcommand{\mdot}{\ifmmode \dot{\mathscr{M}}  \else $\dot{\mathscr{M}}$\ \fi}
\newcommand{\rhb}{\ifmmode R_{\rm BLR}({\rm H\beta})  \else $R_{\rm BLR}({\rm H\beta})$ \ \fi}
\newcommand{\shb}{\ifmmode \sigma_{\rm H\beta} \else $\sigma_{\rm \hb}$\ \fi}
\newcommand{\RL}{\ifmmode R_{\rm BLR}({\rm H\beta}) - L_{\rm 5100} \else $R_{\rm BLR}({\rm H\beta}) - L_{\rm 5100}$ \ \fi}
\newcommand{\ms}{\ifmmode M_{\rm BH}-\sigma_{\ast} \else $M_{\rm BH}-\sigma_{\ast}$\ \fi}
\newcommand{\sm}{\ifmmode \sigma_{\rm H\beta,mean} \else $\sigma_{\rm H\beta,mean}$\ \fi}
\newcommand{\sr}{\ifmmode \sigma_{\rm H\beta,rms} \else $\sigma_{\rm H\beta,rms}$\ \fi}
\newcommand{\fwm}{\ifmmode \rm FWHM_{\rm mean} \else $\rm FWHM_{\rm mean}$\ \fi}
\newcommand{\fwr}{\ifmmode \rm FWHM_{\rm rms} \else $\rm FWHM_{\rm rms}$\ \fi}
\newcommand{\vpfm}{\ifmmode \rm VP_{\rm F, mean} \else $\rm VP_{\rm F, mean}$\ \fi}
\newcommand{\vpsm}{\ifmmode \rm VP_{\rm \sigma, mean} \else $\rm VP_{\rm \sigma, mean}$\ \fi}
\newcommand{\vpfr}{\ifmmode \rm VP_{\rm F, rms} \else $\rm VP_{\rm F, rms}$\ \fi}
\newcommand{\vpsr}{\ifmmode \rm VP_{\rm \sigma, rms} \else $\rm VP_{\rm \sigma, rms}$\ \fi}
\newcommand{\ffm}{\ifmmode f_{\rm F, mean} \else $f_{\rm F, mean}$\ \fi}
\newcommand{\fsm}{\ifmmode f_{\rm \sigma, mean} \else $f_{\rm \sigma, mean}$\ \fi}
\newcommand{\ffr}{\ifmmode f_{\rm F, rms} \else $f_{\rm F, rms}$\ \fi}
\newcommand{\fsr}{\ifmmode f_{\rm \sigma, rms} \else $f_{\rm \sigma, rms}$\ \fi}
\newcommand{\fc}{\ifmmode f_{\rm c} \else $f_{\rm c}$\ \fi}
\newcommand{\dhbm}{\ifmmode D_{\rm H\beta,mean} \else $D_{\rm H\beta,mean}$\ \fi}
\newcommand{\dhbr}{\ifmmode D_{\rm H\beta,rms} \else $D_{\rm H\beta,rms}$\ \fi}

\begin{document}

\title{The supermassive black hole masses of reverberation-mapped active galactic nuclei}

\correspondingauthor{W. -H. Bian}
\email{whbian@njnu.edu.cn}

\author[0000-0001-7552-8895]{Li-Ming Yu}
\affiliation{School of Physics and Technology, Nanjing
Normal University, Nanjing 210046, China}

\author[0000-0002-2121-8960]{Wei-Hao Bian}
\affiliation{School of Physics and Technology, Nanjing
Normal University, Nanjing 210046, China}

\author{Xue-Guang Zhang }
\affiliation{School of Physics and Technology, Nanjing
Normal University, Nanjing 210046, China}

\author{Bi-Xuan Zhao}
\affiliation{School of Physics and Technology, Nanjing
Normal University, Nanjing 210046, China}

\author{Chan Wang}
\affiliation{School of Physics and Technology, Nanjing
Normal University, Nanjing 210046, China}

\author{Xue Ge}
\affiliation{School of Physics and Technology, Nanjing
Normal University, Nanjing 210046, China}

\author{Bing-Qian Zhu}
\affiliation{School of Physics and Technology, Nanjing
Normal University, Nanjing 210046, China}

\author{Yu-Qin Chen}
\affiliation{School of Physics and Technology, Nanjing
Normal University, Nanjing 210046, China}

\begin{abstract}
Using different kinds of velocity tracers derived from the broad \hb profile (in the mean or rms spectrum) 
and the corresponding virial factors $f$, the central supermassive black hole (SMBH) masses (\mbh) are calculated for a compiled sample of 120 reverberation-mapped (RM) AGNs. For its subsample of 
RM AGNs with measured stellar velocity dispersion (\sst), 
the multivariate linear regression technique is used to calibrate the mean value $f$, as well as the variable FWHM-based $f$. 
It is found that, whether excluding the pseudo-bulges or not, \mbh from the \hb line dispersion in the mean spectrum (\sm) has the smallest offset rms with respect to the \ms relation. 
For the total sample excluding SDSS-RM AGNs, 
with respect to \mbh from \sst or that from the \hb line dispersion in the rms spectrum (\sr), it is found that we can obtain \mbh from the \sm with the smallest offset rms of 0.38 dex or 0.23 dex, respectively. It implies that, with respect to the \hb FWHM, we prefer \sm to calculate \mbh from the single-epoch spectrum. Using the FWHM-based $f$, we can improve \mbh calculation from FWHM(\hb) and the mean $f$, with a decreased offset rms from 0.52 dex to 0.39 dex with respect to \mbh from \sst for the subsample of 36 AGNs with \sst. The value of 0.39 dex is almost the same as that from \sm and the mean $f$. 

\end{abstract}



\section{Introduction} \label{sec:intro}

There is good observational evidence that supermassive black holes (SMBHs) exist in nearby galaxies (quiescent or active ), as well as faraway galaxies in universe. Understanding the properties of these SMBHs will clarify their roles in galaxy formation and evolution across the cosmology history, as well as the physics of Active galactic nuclei (AGNs)  \citep[e.g.,][]{bian2002, KH13, N2013}. There are mainly two parameters for SMBHs, i.e., mass (\mbh) and spin, which need to be determined. Combining with the bolometric luminosity, we can investigate the SMBH accretion process and its connection with different components (e.g., hot corona, gas, dust torus, jet) in different classes of AGNs \citep[e.g.,][]{Marconi2004, N2013, N2019, Du16a, Wang2019}.

For more than one hundred of nearby ($<$ 300 Mpc) galaxies, including our Galaxy, SMBH masses have been directly measured through the stellar dynamics, gas dynamics, kinematics of mega-masers, proper motion, or recent direct imaging technique \citep[e.g.,][]{KH13, Sahu2019}. It was suggested that there is a co-evolution of SMBHs and host galaxies \citep[e.g.,][]{KH13}. The quiescent galaxies follow a correlation between the SMBH mass and the bulge or spheroid stellar velocity dispersion (\sst) called the \ms relation  \citep[e.g.,][]{Tr02,Mc11,KH13,Sahu2019},
\begin{equation}
 \label{eq1}
\log \frac{\mbh (\sigma_*)}{10^9 \msun}=\alpha+\beta \log \frac{\sst}{~200~\kms},
\end{equation}
Considering different definition of \sst and different bulge type, there are different values of $\beta$ and  $\alpha$ \citep[e.g.,][and reference therein]{KH13, HK14, Ba17}.  
It was found that local AGNs with directly measured \mbh follow the \ms relation for the quiescent galaxies, indicating that galaxies with and without an AGN follow a single relation \citep[e.g.,][]{Sahu2019}.

For faraway AGNs, because of a limit of space resolution and  outshining their hosts , it is too difficult to weigh the SMBH masses through above stellar kinematics or gaseous dynamics method \citep[e.g.,][]{Sahu2019}. AGNs can be classified into type 1 or type 2 AGNs, depending on whether the broad-line regions (BLRs) can be viewed directly \citep{N2013}. For type 1 AGN, the BLR clouds can be used as a probe of the gravitational potential of the SMBH of which virial mass can be derived as follows \citep[e.g.,][]{Pe04}:
 \begin{equation}
 \label{eq2}
\mbh=f\times \frac{R_{\rm BLR}~(\Delta V)^2}{G} \equiv f\times \rm VP .
\end{equation}
where  $\Delta V$ is the velocity of the BLR clouds under the virialization assumption, $R_{\rm BLR}$ is the distance from the SMBH to the BLRs,  $f$ is a virial factor, and $G$ is the gravitational constant. VP is the so-called virial product, ${\rm VP} = R_{\rm BLR}~(\Delta V)^2/G$. The virial factor $f$  is used to characterize the kinematics, geometry, inclination of the BLR clouds \citep[e.g.,][]{Co06, Yu19}. $R_{\rm BLR}$  can be estimated from the reverberation mapping (RM) method \citep[e.g.,][]{BM82,Pe93} and from the empirical \RL relation ($L_{\rm 5100}$ is the 5100 \AA\ luminosity) for the \hb broad line derived based on the RM AGNs \cite[e.g.,][]{Ka00, bian2004, Ka05, bian2008, Sh11, Be13, KE15, Du18, Wang2019, Du19, Yu20}. The velocity of the BLR clouds $\Delta V$ is  usually traced by the Full-width at half-maximum (FWHM) or the line dispersion ($\sigma_{\rm \hb}$) of the broad \hb emission line measured from the mean or rms spectrum \citep{Pe04, Co06}. The line dispersion from the rms spectrum \sr was believed to be the best tracer of $\Delta V$ in the calculation of SMBH masses \citep{Pe04}. The complex \hb profile was suggested to be created by different components of spiral-in BLR clouds from tidally disrupted dusty clumps, i.e.,  inflow, ouflow, and circularized gas \citep{Wang2017}.  How to derive $\Delta V$ from the complex profile of the broad line is still an open question, especially for other high-ionized broad lines, such as \civ\ \citep[e.g.,][]{Ge19}.

In order to weigh the SMBH masses in RM AGNs through equation (\ref{eq2}), the factor $f$ is a key quantity, which is usually done through RM AGNs following the \ms relation for quiescent galaxies \citep[e.g.,][]{On04, Gr13, Woo13, HK14, Woo15, Ba17, Yu19, WS2019} or other independent methods to derive the SMBH masses in RM AGN \citep{Sturm2018, Williams2018, Mejia2018}. Assuming that AGNs follow the \ms relation defined by the quiescent galaxies, we can derive \mbh from \sst. For other independent methods to derive \mbh in AGNs, we can also use them to calibrate $f$. Using the \ms relation for quiescent galaxies, the mean values of $f$ are achieved for different $\Delta V$ tracers for RM AGNs with measured \sst \citep[e.g.,][]{HK14}. For each RM AGN with \mbh weighed from the measured \sst, we can calculate $f$ for individual AGN \citep{Yu19}. We found a wide distribution of $f$ and did the calibration of variable $f$, i.e., $f \propto \rm FWHM(\hb)^{-1.11}$ when FWHM(H$\beta$) is used as the tracer of $\Delta V$ in equation (\ref{eq2}) \citep{Yu19}. Our result is consistent with results from the accretion disk (AD) model to fit the AD spectra of AGNs \citep{Mejia2018} and the BLRs dynamical model to fit simultaneously the AGNs continuum/\hb light curves and \hb line profiles \citep[e.g.,][]{Li2018, Pancoast2018, Williams2018}. For other un-RM AGNs yet, the empirical \RL relation was investigated \citep[e.g.,][]{Ka00, Ka05, Be13, Du19, Yu20}. The virial factor $f$ and the empirical \RL relation provide the foundation of \mbh calculation from the single-epoch spectrum of a type I AGN from large spectral surveys \citep[e.g.,][]{bian2004, VP06, Sh11, Woo15}.  


In this paper, using different kinds of velocity tracers from the broad \hb profile in the mean or rms spectrum and the corresponding virial factors, the SMBH masses are investigated for a compiled sample of 120 RM AGNs and its subsample of 36 RM AGNs with measured \sst. Section 2 describes our sample. Section 3 is the calibration of constant or variable $f$ for the subsample of 36 RM AGNs using the multivariate linear regression technique. Section 4 is the comparison of the SMBH masses from different methods for the sample of 120 RM AGNs. Section 5 summarizes our results.
All of the cosmological calculations in this paper assume $\Omega_{\Lambda}=0.68$, $\Omega_{\rm M}=0.32$, and $H_{0}=67~ \kms {\rm Mpc}^{-1}$ \citep{Plank2014}.

\section{The Sample of RM AGNs}\label{sec:samp}
Up to now, there are about 120 AGNs complied with measured the \hb time lags from the RM method to investigate the extended \RL relation in our previous work  \citep{Yu20}. We had divided them into three subsamples, i.e., BenzSample, SEAMBH, and SDSS-RM \citep[][and references therein]{Yu20}. The first subsample has 25 AGNs as super-Eddington accretor ( $\mdot \geq 3$) presented by SEAMBH collaboration \citep[hereafter SEAMBHs;][]{Du15, Du16b, Du18}. The second subsample has 39 AGNs summarized by \cite{Be13} and 12 other sources published recently \citep[hereafter BentzSample;][]{Ba15, Be16a, Be16b, Fa17, Williams2018}. The third subsample contains 44 high-$z$ AGNs ($z \sim 0.1-1.0$) from the Sloan Digital Sky Survey (SDSS) RM Project which was done by the fibre spectrum \citep[hereafter SDSS-RM;][]{Gr17}. 

In order to calculate the virial \mbh, we need to know the velocity of the BLR clouds. Using the broad \hb emission line, there are four kinds of the BLRs velocity tracers from the mean and rms spectrum, i.e., \fwm, \sm, \fwr, \sr, respectively \citep[e.g.,][]{HK14, Yu19, Yu20}. The contributions from the narrow-component and instrumental broadening in the \hb emission line were subtracted to measure the line width for the subsample of SDSS-RM and SEAMBH \cite[e.g.,][]{Sh15, Du18}. 
For \fwm and \sm from the mean spectrum, the data were presented in our previous paper \citep[][and the reference therein]{Yu20}. For \sr or \fwr from the rms spectrum, we collect them for 12 out of 25 AGNs in the subsampel of SEAMBH \citep[e.g.,][]{Du18}, 50 out of 51 AGNs (except  Mrk 1511) in the subsample of BentzSample \citep[e.g.,][]{Co06,Ba13,Fa17, Williams2018}, and all of 44 AGNs in the subsample of SDSS-RM \citep{Gr17}. There are 106 out of 120 AGNs with collected data from the rms spectra. Properties about our sample of 120 RM AGNs are shown in Table \ref{tab1} \citep[see also][]{Yu20}. For the total sample, there are 37 narrow-line Seyfert 1 galaxies (NLS1s) with \fwm(\hb) $< 2000 \kms$ \citep[e.g.,][]{bian2004}.

In equation \ref{eq2}, we need to calibrate the virial factor $f$. Our subsample for $f$ calibration consists of 36 low redshift broad-line AGNs ($z<0.1$
except PG 1617+175) with both measured \hb lags and reliable \sst, which allows us to calibrate the factor $f$ based on the \ms relation \citep[e.g.,][]{Yu19}.
32 out of these 36 RM AGNs are selected from \cite{HK14}, who had imaged these objects and classified them into three bulge types: elliptical, classical bulges (CB) and pseudobulge (PB). Bulge is considered as PB with a S\'ersic index $n < 2$ or the bulge-to-total (B/T) luminosity ratio not larger than 0.2 \citep{HK14}. 
Beyond the sample of \cite{HK14}, there are four additional objects. MCG+06-26-012 ($\sst=112~\kms$) is classified with pseudobulge by \cite{Wang2014}. Its RM result is from \cite{Du15} and its stellar velocity dispersion is adopted from \cite{Woo15}. NGC 5273 ($\sst=74~\kms$) is a early type galaxy that reverberation mapped by \cite{Be14}. UGC 06728 ($\sst=52~ \kms$, B/T $\sim$ 0.2) and MCG-06-30-15 ($\sst=109\pm 9 ~\kms$, n=1.9) are two late type galaxies with S\'ersic bulges  \citep{Be16a,Be16b}. 
Therefore, our subsample consists with 7 ellipticals, 10 classical bulges and 19 pseudobulges. There are 9 NLS1s out of 36 RM AGNs with measured \sst. For 12 AGNs and 32 quiescent galaxies with spatial-resolved \sst data, \cite{Ba17} presented the spatial-resolved \sst. On average, it is smaller by 13 $\sim \kms$ than from a single spectrum integrated within the effective radius $r_e$ containing half light from the whole galaxy/bulge \citep{Ca13}.  
We adopted the most widely available \sst data from  \cite{KH13}\citep[see also][]{Gu09, HK14}.
Properties about our subsample of 36 low-z AGNs with measured \sst are shown in Table \ref{tab2}.

For a high-$z$ sample of 44 AGNs ( $z \sim 0.1 - 1.0$) from SDSS-RM Project \citep{Gr17} with measured \hb/\ha lags (44 \hb lag, 18 \ha lags), there are 26 of 44 SDSS-RM AGNs with measured \hb lag and \sst by \cite{Sh15}. Four kinds of the BLRs velocity from the \hb profile for these 26 SDSS-RM AGNs are given in Table 2 in \cite{Yu19}.


For objects mapped many times, we use the square of measurement error as weight to calculate the weighted average. The error of the compiled data is calculated from the weighted measurement error and the weighted standard deviation  \citep{Du15,Du19,Yu20}.

\section{The calibrating the virial factor from 36 RM AGNs with measured \sst}\label{sec:calif}
In our previous paper \citep{Yu19}, we have calculated the virial factor $f$ for each object in 34 AGNs with measured \sst through the SMBH mass estimated from \sst (assuming the \ms relation) and the virial product from the \hb time lag and the BLRs velocity. For four tracers of the velocity of the BLRs, there are four kinds of the factor $f$ for every RM AGNs. We found significant correlations between the FWHM-based $f$ and the \hb FWHM, the line dispersion. In this work, considering the intrinsic scatter in the \ms relation, we use the multivariate regression analysis technique to derive the mean and variable factor of $f$, and then investigate the SMBH masses in RM AGNs \citep{Me03,HK14, Yu20}. 

\subsection{Multivariate liner regression technique}
We briefly described the multivariate regression analysis technique as follows, which is used in our investigation of the extended \RL relation \cite[for detail in ][]{Yu20}. For the best linear fit in the form: $ y = \Sigma_{j}\beta_{j}x_{j} + \alpha$, the estimator of $\chi^2$ is used to find the best values for these parameters \citep[e.g.,][]{Me03, On04, HK14, Woo15, Yu20}:
 \begin{equation}
\chi^2 = \Sigma_{i} \frac{(y_{i} - \Sigma_{j}\beta_{j}x_{ij}-\alpha)^2}{\sigma_{\rm int}^{2}+\sigma_{y_{i}}^{2}+\Sigma_{j}(\beta_{j}\sigma_{x_{ij}})^2},
 \label{eq3}
\end{equation}
where $y_{i}$ is the dependent variable. $x_{ij}$ are the independent variables. $\sigma_{y_{i}}$, $\sigma_{x_{ij}}$ are the uncertainties of $y_{i}$, $x_{ij}$, $\alpha$ is the zero intercept, $\beta_{j}$ are the regression coefficients and  $\sigma_{\rm int}$ is the intrinsic scatter. 
For a given set of $\beta_{j}$ , solving the equation $\frac{\partial\chi^2}{\partial\alpha}$ = 0, the optimal
value $\alpha_{\rm min}$ is
\begin{equation}
\label{eq4}
\alpha_{\rm min}=\frac{\Sigma_{i}\frac{y_{i}-\Sigma_{j}\beta_{j}x_{ij}}{\sigma_{\rm int}^{2}+\sigma_{y_{i}}^{2}+\Sigma_{j}(\beta_{j}\sigma_{x_{ij}})^2}}{\Sigma_{i}
(\sigma_{\rm int}^{2}+\sigma_{y_{i}}^{2}+\Sigma_{j}(\beta_{j}\sigma_{x_{ij}})^2)^{-1}},
\end{equation}
The value of $\sigma_{\rm int}$ can be derived by iteration with $\chi^{2}_{r}=1$  where $\chi^2_{r}=\chi^2/N_{dof}$, $N_{dof}$ is the number of degree of freedom \citep{Bam06, Bed06, Pa12, Yu20}. 
\begin{equation}
\label{eq5}
\sigma^2_{\rm int,j+1}=\sigma^2_{\rm int,j} \chi_{r}^{4/3},
\end{equation}


The error bars of $\alpha$ , $\beta_{j}$ and $\sigma_{\rm int}$ are estimated by the bootstrap method. We re-sampled the data for 100 times and repeated the fitting procedure to estimate the uncertainties of these fitting parameters.

\subsection{The mean values of $f$ based on different velocity tracers}
\begin{figure*}
\includegraphics[angle=-90,width=3.4in]{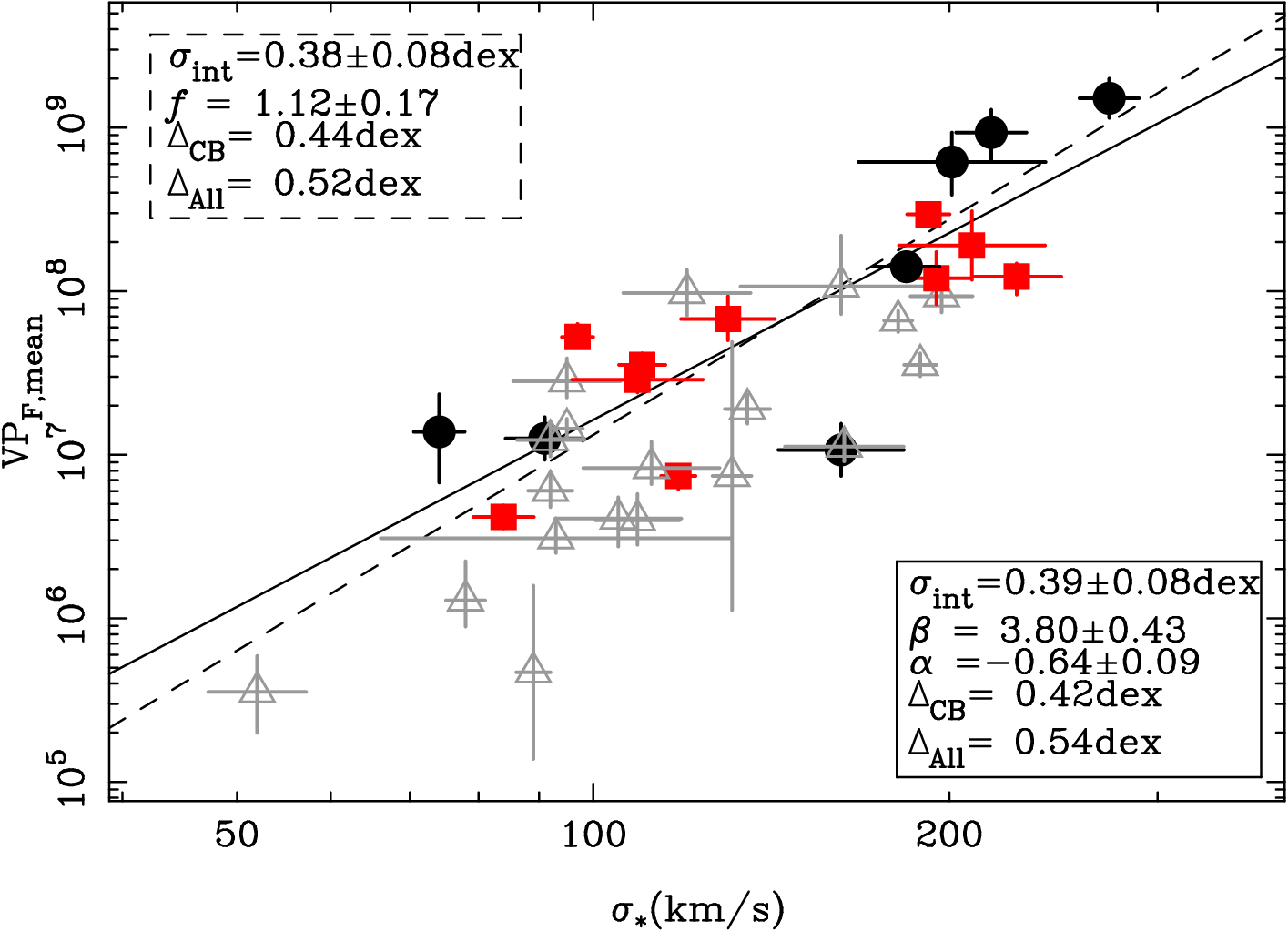}\hfill
\includegraphics[angle=-90,width=3.4in]{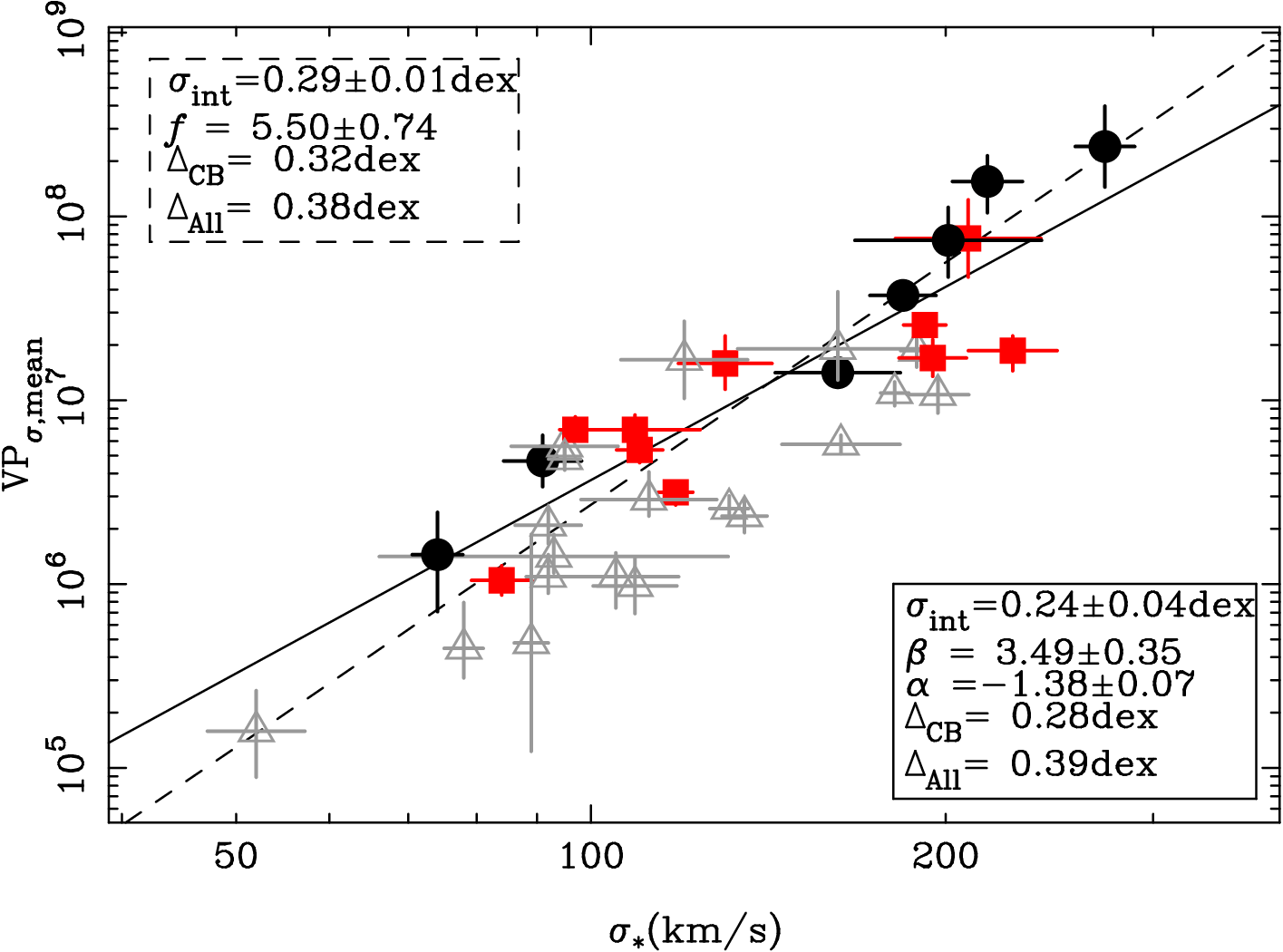}\vfill
\includegraphics[angle=-90,width=3.4in]{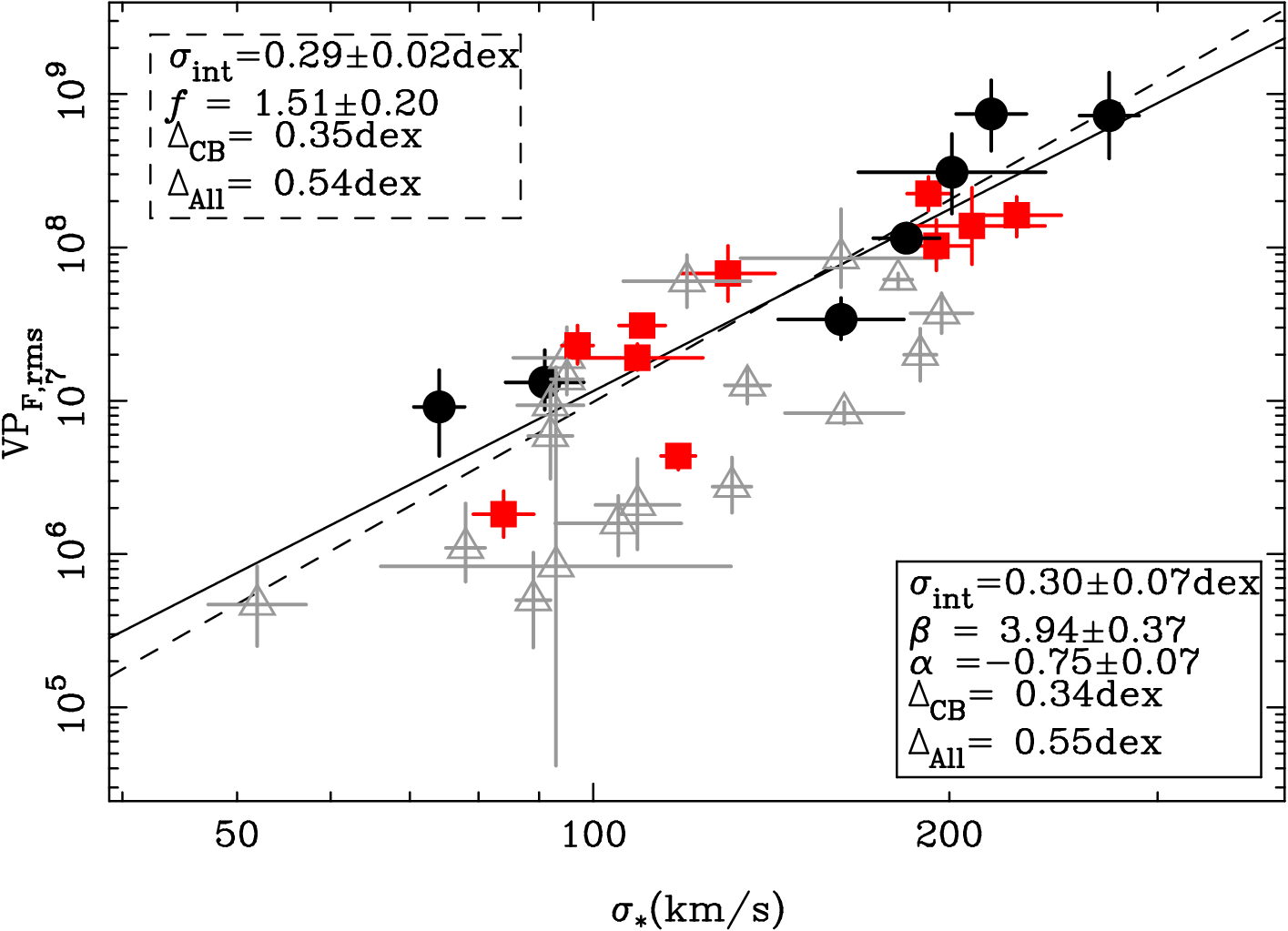}\hfill
\includegraphics[angle=-90,width=3.4in]{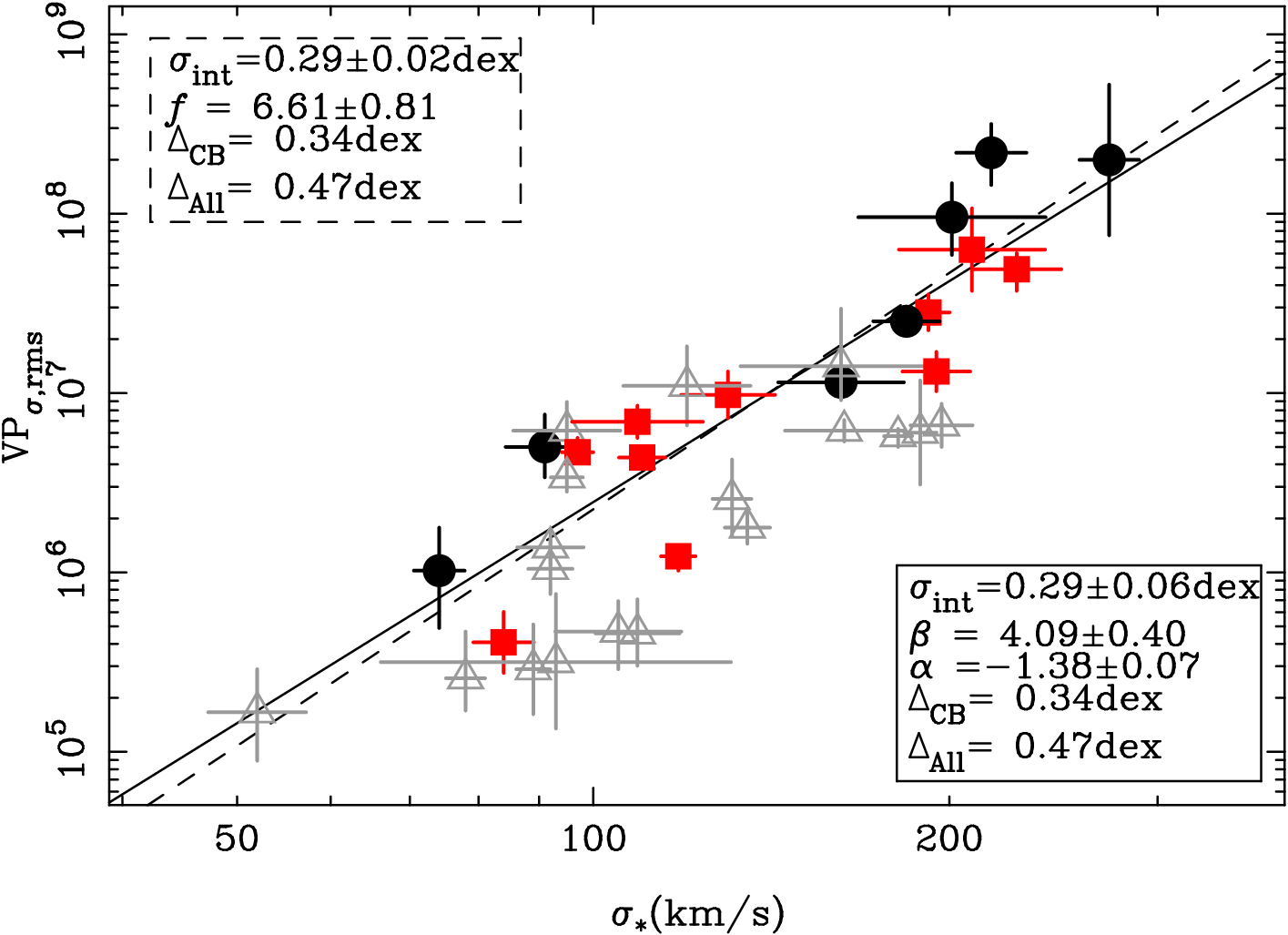}\hfill

\caption{Four kinds of VPs versus \sst. Top left: \vpfm versus \sst. Top right: \vpsm versus \sst. Bottom left: \vpfr versus \sst. Bottom right : \vpsr versus \sst. Black circles denote ellipticals, red squares denote classical bulges, and gray triangles represent pseudo-bulges. Our best-fit line for ellipticals and classical bulges (excluding PBs) is shown as the black solid line. The detail about the best-fit line is shown in the bottom right corner ($\sigma_{\rm int}$, the slope, the intercept, the offset rms for ellipticals and classical bulges, the offset rms for all AGNs). Fixing the slope ($\beta$ = 4.38) same as the \ms relation for quiescent galaxies \citep{KH13}, we can fit the intercept to obtain the virial factor $f$ for ellipticals and classical bulges. The dashed line is our best fit with a fixed slope ($\beta$ = 4.38). The detail is presented in the upper left corner ($\sigma_{\rm int}$, the mean virial factor, the offset rms for ellipticals and classical bulges, the offset rms for all AGNs). }
\label{fig1}
\end{figure*}

It was shown that only the galaxies with classical bulges and ellipticals obey a tight \ms relation, but not PBs \citep{KH13}. \cite{HK14} assumed that RM AGNs with classical bulges and ellipticals intrinsically obey the same \ms relation as the local inactive galaxies, but for PB with the same \sst, the \sst - based \mbh is smaller and scaled by a factor of 3.80 \citep[see also][]{Yu19}. With spatially resolved kinematics from integral-field spectroscopy, it was found that the rotational broadening of the spectrum typically would flatten the slope of the \ms relation, $\beta=4.76\pm 0.60$ for 32 quiescent galaxies, 
$3.90\pm 0.93$ for 16 AGNs, respectively \citep{Ba17}. Different slope of the \ms relation in quiescent galaxies or AGNs is possibly due to different methods to derived \sst, the measurement of $r_e$, the effect of rotational broadening (galaxy morphologies and disk inclinations) \cite[e.g.,][and reference therein]{Ba17}. Rotational broadening typically shallows the slope of the \ms relation for AGNs \citep{Ba17}. The slope difference also maybe an artifact of the sample selection bias \citep{Woo13, Sh16}. 
Using FWHM-based $f$ from the mean spectrum, it was found that SDSS-RM AGNs followed the \ms relation and showed a larger scatter of \mbh than that for low-z RM AGNs \citep{Yu19}. It implied the possibility of evolution of the \ms relation for high-z AGNs or the large systematic uncertainties in \mbh and \sst. In this work, like as \cite{HK14}, we do $f$ calibration assuming that these low-z subsample of 17 RM AGNs with E/CBs intrinsically obey the same \ms relation as the local quiescent galaxies excluding the pseudo-bulges, i.e., $\beta=4.38\pm 0.29$, $\alpha=-0.51 \pm 0.05$ \citep{KH13}. In this section, all the best fits are done for low-$z$ sample excluding PBs, unless otherwise stated.

It was also suggested that AGNs with PBs follow the \ms relation for the quiescent galaxies \citep[e.g.,][]{Gr12, Woo15}. 
\cite{Sh15} showed no evidence of \ms evolution up to $z\sim 1$ for SDSS-RM AGNs. 
\cite{Li2020} also suggested no strong evidence of evolution in the relation between $M_{\rm BH}$ and the stellar mass  $M_{\star,bulge}$ to $z \sim$ 0.6. Considering no $z$ evolution in this relationship suggested by them, we also include these 26 high-z SDSS-RM AGNs to calibrate $f$, as well as PBs, which are shown in the Appendix (see Fig. \ref{fig1-1}, \ref{fig2-1}, Table \ref{tab3}).

To determine the virial factor, we use the $\chi^2$ estimator as discussed in Section 3.1, where $y$ in equation (\ref{eq3}) is $\log \rm VP$, $x_j$ is $\log \sst$:
\begin{equation}
\label{eq6}
\chi^2=\Sigma_{i}\frac{(\log f+\log({\rm VP}_{i})-\beta\log\sigma_{\ast,i}-\alpha)^2}{\sigma_{{\rm VP}_{i}}^2+(\beta \sigma_{\log\sigma_{\ast,i}})^2+\sigma_{\rm int}^2}.
\end{equation}
Fixing the slope $\beta$, for a giving $\sigma_{\rm int}$, we can use the equation (\ref{eq4}) solve the
value of the zero intercept, $\log f - \alpha$, analytically. 
For 17 low-z RM-AGNs with E/CBs (excluding PBs and high-$z$ SDSS-RM AGNs), for a fixed value of the slope of $\beta = 4.38$ \citep{KH13, HK14}, we obtain the value of $\log f - \alpha$, $\sigma_{\rm int}$ for $\chi_{r}^2=1$ in equation \ref{eq6}. Using $\alpha = -0.51$, we obtain the mean values of $f$ for different kinds of VPs. 
The intrinsic scatter $\sigma_{\rm int}$, virial factor $f$, the offset rms (with respect to the \ms relation) including PBs or not are shown in the upper left corner of each panel in Fig. \ref{fig1}. For four kinds of velocity tracer $\Delta V$, the best-fit of four kinds of virial  factor are $\ffm = 1.12 \pm 0.17$, $\fsm = 5.50 \pm 0.74$, $\ffr = 1.51 \pm 0.20$ and $\fsr = 6.61 \pm 0.81$ (see Table \ref{tab3}), which are consistent well with our previous results by \cite{Yu19}, i.e., 1.20, 5.75, 1.48, 6.03, respectively. They are also  consistent with other works \citep{Gr13, Woo13, HK14, Woo15}. For an RM sample of 29 AGNs with measured \sst, \cite{Woo15} also found that $\beta=5.04\pm 0.28, \ffm=1.12$ by jointly fitting the \ms relation using 84 quiescent galaxies and 29 RM AGNs with measured \sst. 

Fig. \ref{fig1} shows the relations between VP and \sst for different VP, and the dashed line is for the case of fixed slope of $\beta=4.38$ . We find that \vpsm is the best VP to calculate the SMBH masses, the intrinsic scatter is 0.29 dex 
and the deviation is the smallest. The offset rms (with respect to \ms relation) is also smallest, 0.32 dex or 0.38 dex whether excluding PBs or not (see the value of $\Delta$ in left corner of each panel in Fig. \ref{fig1}). Although the line dispersion in the rms spectrum was suggested to be a best tracer of $\Delta V$ from AGNs with multiple mappings \citep{Pe04}, the smallest offset rms using \sm implies that we can also use \sm to trace the velocity of the BLR clouds \citep[see also][]{Woo15}. 

Adopting different \ms relations, for $\beta=5.32\pm 0.63$, $\alpha=-0.45\pm 0.09$ \citep{Ba17}, the mean values of 4 kinds of $f$ become smaller ($\ffm=0.93\pm 0.22, \ffr=1.23\pm 0.24, \fsm=4.57\pm 0.89, \fsr=5.50\pm 0.87$). With $\beta=4.76\pm 0.60$,$\alpha=-0.34\pm 0.09$ \citep{Ba17}, the mean values of 4 kinds of $f$ become larger ($\ffm=1.48\pm 0.24, \ffr=1.95\pm 0.27, \fsm=7.08\pm 1.31, \fsr=8.71\pm 1.58$). Therefore, the calibration of mean $f$ for the virial \mbh has a dependence on the \ms relation. 

We also fit the data when $\beta$ and $\alpha$ are unconstrained. Because the $\alpha$ and $f$ in equation (\ref{eq6}) are degenerate, we can not get $f$. The form of $\chi^2$ is:
\begin{equation}
\label{eq7}
\chi^2=\Sigma_{i}\frac{(\log({\rm VP}_{i})-\beta\log\sigma_{\ast,i}-\alpha)^2}{\sigma_{{\rm VP}_{i}}^2+(\beta\sigma_{\log\sigma_{\ast,i}})^2+\sigma_{\rm int}^2}.
\end{equation}
Excluding the pseudo-bulges in the fitting, the best-fits of the four kinds of VP versus \sst are demonstrated as solid lines in Fig. \ref{fig1}. We also present the intrinsic scatter, slope $\beta$, intercept $\alpha$, the offset rms for ellipticals and classical bulges, the offset rms for all AGNs in the bottom right corner of each panel. All the slopes are less than 4.38 suggested by \cite{KH13}. The slope of the relation of \vpsr versus \sst is consistent with 4.38 in $1\sigma$ level, the slopes of others are consistent with 4.38 in $2\sigma$ level. The slope of the relation of \vpsm versus \sst is $3.49\pm 0.35$, the smallest among four kinds of VP versus \sst. It is consistent with the result of $3.46\pm 0.61$ for \vpsm by \cite{Woo13} for a RM sample of 25 AGNs with measured \sst \citep[see also][]{Woo10}, and with the result of $3.90\pm 0.93$ for 16 AGNs by \cite{Ba17} .
For a sightly larger RM sample of 29 AGNs with measured \sst, adopting $\ffm=1.12$,  \cite{Woo15} found that the slope $\beta$ is $4.32\pm  0.59$ using \vpfm, which is flatter than $\beta=5.04\pm 0.28$ in their jointly fitting the \ms relation using the quiescent galaxies and RM AGNs. Our flatter slope is possibly due to the rotational broadening effect in \sst, or an artifact of the sample selection bias \citep[e.g.][]{Ba17}.

Above fitting results are obtained for low-z RM AGNs excluding PBs, but we still demonstrate PBs as gray triangles in Fig.\ref{fig1}. PBs are flowing the solid fitting line for ellipticals and CBs, which are consistent with others \citep{Gr12,Woo15}. It implicates that AGN with PBs may follow the same \ms relation as quiescent galaxies. In Fig. \ref{fig1-1} in the Appendix section, including high-z SDSS-RM AGNs and PBs, we also find similar results and present in Table \ref{tab3}. In this paper, we prefer the results from the fitting for 17 low-$z$ RM-AGNs with E/CBs.

\subsection{The FWHM-based $f_{c}$}
\begin{figure}
\includegraphics[angle=-90,width=3.0in]{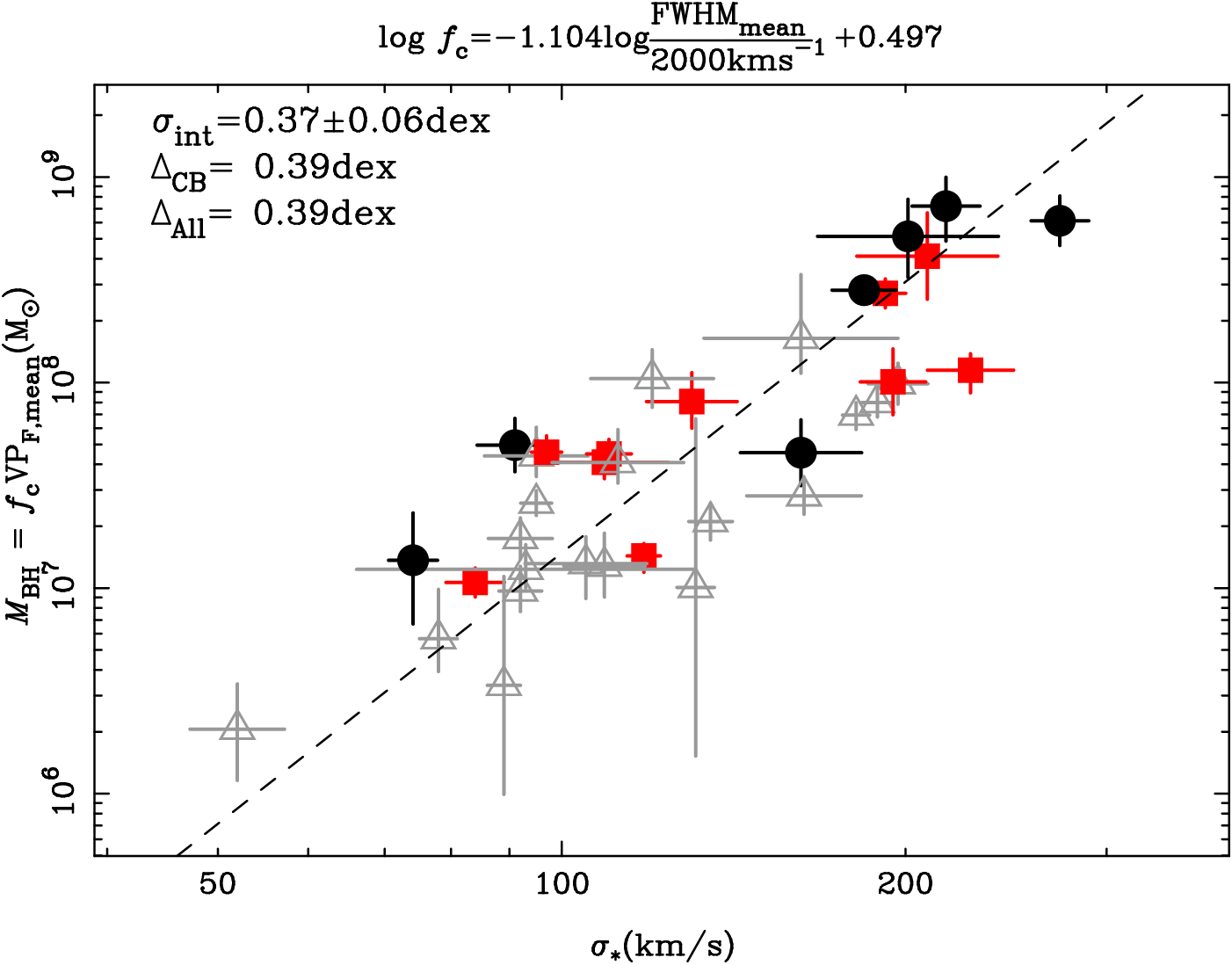}
\includegraphics[angle=-90,width=3.0in]{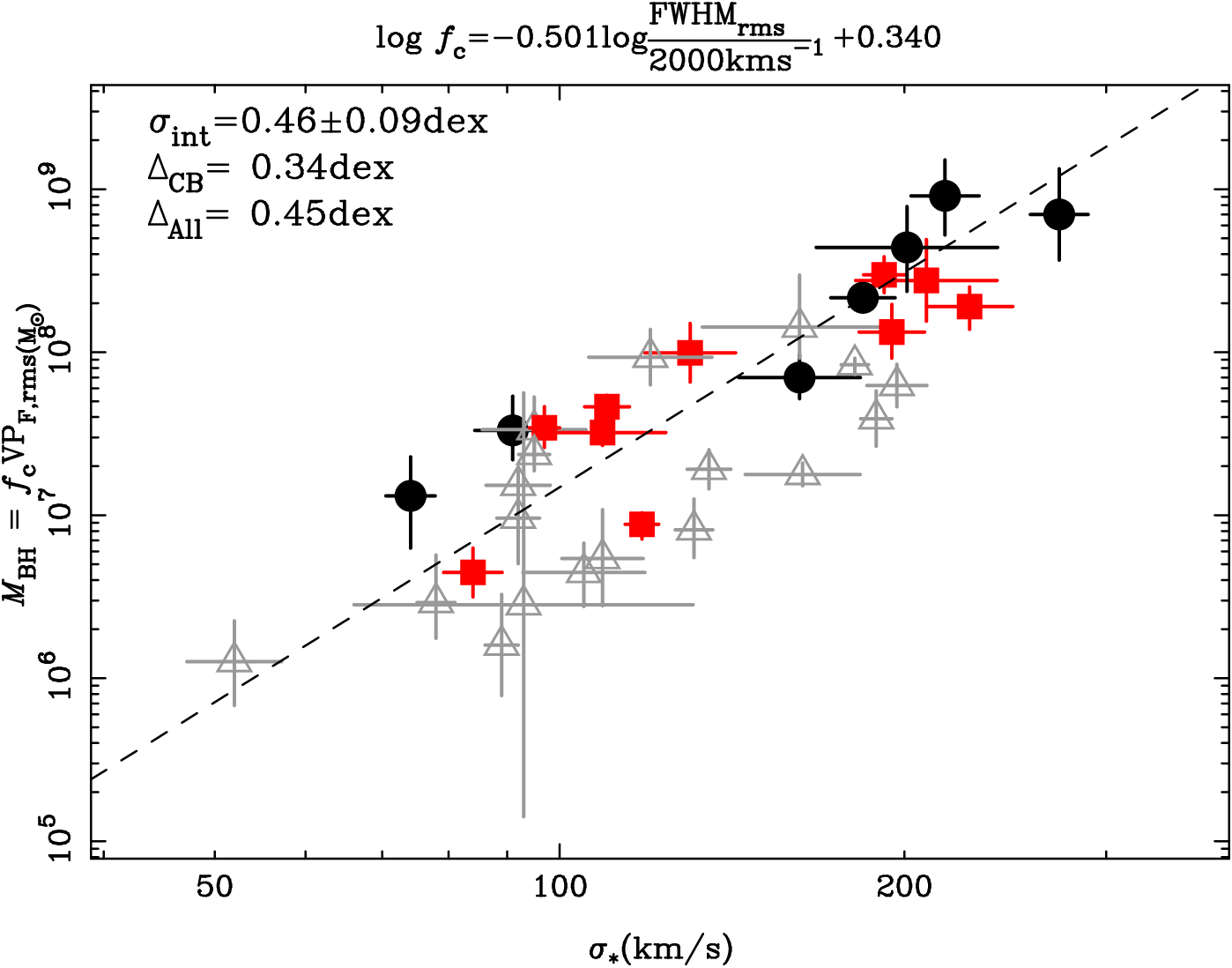}\hfill
\caption{The \ms relation. Top: \mbh are calculate from \vpfm, and we show the virial factor corrected by \fwm on the top of the figure. Bottom: \mbh are calculate from \vpfr, and we also show the virial factor corrected by \fwr on the top of the figure. The symbols are the same as Fig. \ref{fig1}. The dashed line is the \ms relation for quiescent galaxies \citep{KH13}}.
\label{fig2}
\end{figure}

Significant correlations between the FWHM-based $f$ and the \hb FWHM, the line dispersion have been reported in our previous paper \citep{Yu19}.
Considering the variable $f$ in stead of the constant $f$ alike in above subsection, we assume a linear function between $f$ and \fwm,
\begin{equation}
\label{eq8}
\log \fc = \alpha_{1} + \beta_{1}\log\frac{\fwm}{2000~\kms}.
\end{equation}
To determine $\alpha_1$ and $\beta_1$ in the above equation (\ref{eq8}), we use the $\chi^2$ estimator discussed in Section 3.1:
\begin{equation}
\label{eq9}
\chi^2=\Sigma_{i}\frac{(\log({\rm VP}_{i})+\alpha_{1}+\beta_{1}\log{\fwm}-\beta\log \sigma_{\ast,i}-\alpha)^2}{\sigma_{{\rm VP}_{i}}^2+(\beta_{1}\sigma_{\rm FWHM_{\rm mean,i}})^2+(\beta\sigma_{\log\sigma_{\ast,i}})^2+\sigma_{\rm int}^2}.
\end{equation}
We fix $\beta=4.38$, and $\alpha=-0.51$ \citep{KH13, HK14}. Considering the intrinsic scatter in the \ms relation for AGNs, for a subsample  of 17 AGNs with ellipticals or classical bulges, the best fit is 
\begin{equation}
\label{eq10}
\fc = -(1.10 \pm 0.40)\log\frac{\fwm}{2000 ~\kms} + (0.50 \pm 0.11).
\end{equation}
The best-fitting results of $\sigma_{\rm int}=0.37\pm 0.06$, $\beta_1=-1.10\pm 0.40$ and $\alpha_1=0.50\pm 0.11$ with a scatter of 0.39 dex (see top panel in Fig. \ref{fig2}, Table \ref{tab3}) are consistent with our previous results, i.e., $\beta_1=-0.80\pm 0.45$ and $\alpha_1=0.36\pm 0.19$, from the $f$ calculation for each AGN with measured \sst \citep{Yu19}. \cite{Mejia2018} estimates \mbh through fitting the SD spectra of 37 AGNs ($z\sim 1.5$), observed using the ESO X-Shooter spectrograph which provides simultaneous, very wide wavelength coverage of the AD emission. Their spectral fitting model is the standard, geometrically thin, optically thick AD model including  general relativistic and disc atmosphere corrections. They also suggested a FWHM-based $f$ and the slope is $-1.17\pm 0.11$ for AGNs at $z \sim 1.5$, which is well consistent very well with ours, i.e., $-1.10\pm 0.40$. Our larger error of the slope is due to smaller number of our sample (17 vs. 37).

Adopting different slope of $\beta=5.32\pm 0.63$ 
for the \ms relation \citep{Ba17}, the slope $\beta_1$ for the FWHM-based $f$ (equation \ref{eq8}) is -0.87 but with a larger intrinsic scatter of 0.48 dex. 
Adopting $\beta=4.76\pm 0.60$ 
\citep{Ba17}, the slope $\beta_1$ is -1.01 but with a large intrinsic scatter of 0.41 dex. 

We also use \fwr instead of \fwm in equation (\ref{eq8}), and find $\sigma_{\rm int}=0.46$, $\beta_1=-0.50\pm 0.25$ and $\alpha_1=0.34\pm 0.11$ (see bottom panel in Fig. \ref{fig2}). It is smaller than $\beta_1=-1.29\pm 0.38$ \citep{Yu19}, which was derived by the bivariate correlated errors
and scatter method \citep[BCES;][]{AB96}. Using FWHM-based $f$, the offset rms (with respect to the \ms relation) is decreased from 0.44 dex to 0.39 dex based on \fwm, and decreased slightly from 0.35 dex to 0.34 dex based on \fwr (see left two panels in Fig. \ref{fig1} and Fig. \ref{fig2}). It shows the decrease of the offset rms using \fwr-based $f$ is not more significant than using \fwm-based $f$. Including high-z SDSS-RM AGNs and/or PBs, we also find similar results for FWHM-based $f$ (see Fig. \ref{fig2-1}, Table \ref{tab3}). The FWHM-based variable $f$ suggests the non-virial \mbh calculated from the \hb FWHM.

\section{The SMBH masses of the 120 RM AGNs}\label{sec:mass}
\subsection{The SMBH mass from \sst}
Using the subsample of 36 RM AGNs with measured \sst, we do the calibration of constant $f$ and FWHM-based $f$ in the above section. With our derived $f$ and the velocity of the BLR clouds, we calculate the SMBH masses through equation (\ref{eq2}). We also calculate the mass through the \ms relation for the calibration subsample of 36 AGNs with measured \sst \citep[the slope $\beta = 4.38$,][]{KH13}, as well as for 30 high-$z$ SDSS-RM AGNs with measured \sst \citep{Sh15}.

\begin{figure}
\includegraphics[angle=-90,width=2.8in]{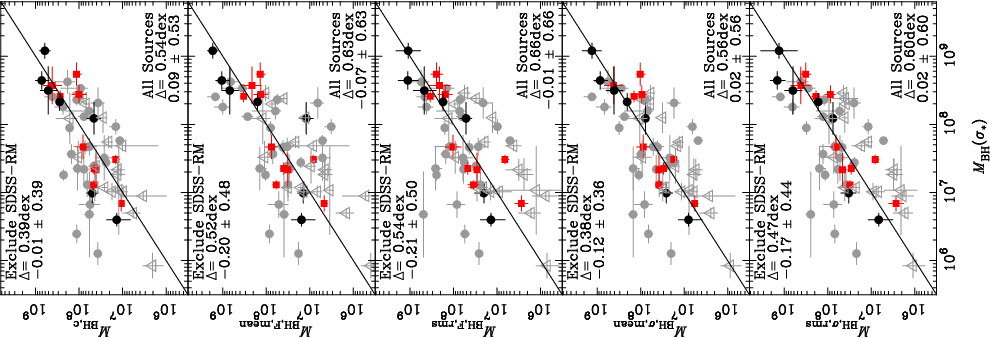}\hfill
\caption{Comparison between the \mbh calculated from \hb line width and the \mbh calculated from \sst. From top to bottom are $M_{\rm BH,c}$, $M_{\rm BH,F,mean}$, $M_{\rm BH,F,rms}$, $M_{\rm BH,\sigma,mean}$ and $M_{\rm BH,\sigma,rms}$ versus \mbh(\sst) respectively. The symbols are the same as Fig. \ref{fig1}.
Gray circles represent the SDSS-RM sources. The solid line in each panel is the 1:1 line. The offset rms (with respect to the solid line), the mean value and its standard deviation of offset for all sources whether excluding SDSS-RM AGNs or not are demonstrated in the left/right corner.  }
\label{fig3}
\end{figure}

Fig. \ref{fig3} shows the comparison of the SMBH masses for these 36 RM AGNs from \sst with that from different velocity tracers and corresponding $f$, as well as for 30 high-$z$ SDSS-RM AGNs with measured \sst (gray circles) \citep{Sh15, Gr17}. Considering large scatter when including SDSS-RM AGNs and possibility of the evolution of the \ms relation for high-$z$ AGNs \citep{Yu19}, we just show them in Figs \ref{fig3}, \ref{fig4} as gray circles and neglect them in our discussion except as otherwise noted. The mass from \sm has the smallest offset rms in the relation with the mass from the \ms relation, 0.38 dex. Using FWHM-based $f$, the offset rms is decreased from 0.52 dex to 0.39 dex, very close to 0.38 dex for the case of \sm (top two panels in Fig. \ref{fig3}). It is consistent with our result in \cite{Yu19}. We also find that the offset rms is 0.47 dex for \sr (the bottom panel in Fig. \ref{fig3}), which is larger than that for \sm (0.38 dex) or FWHM-based $f$ (0.39 dex). Including SDSS-RM sources, the scatter is obvious larger in Fig. \ref{fig3} with an increasing offset rms of 0.55-0.66 dex (legend in right corners in each panel in Fig. \ref{fig3}). The case of FWHM-based $f$ has the smallest offset rms (0.55 dex) and the next is for the case of \sm (0.56 dex).

\subsection{The SMBH mass from the  \hb line}
\begin{figure}
\includegraphics[angle=-90,width=3.0in]{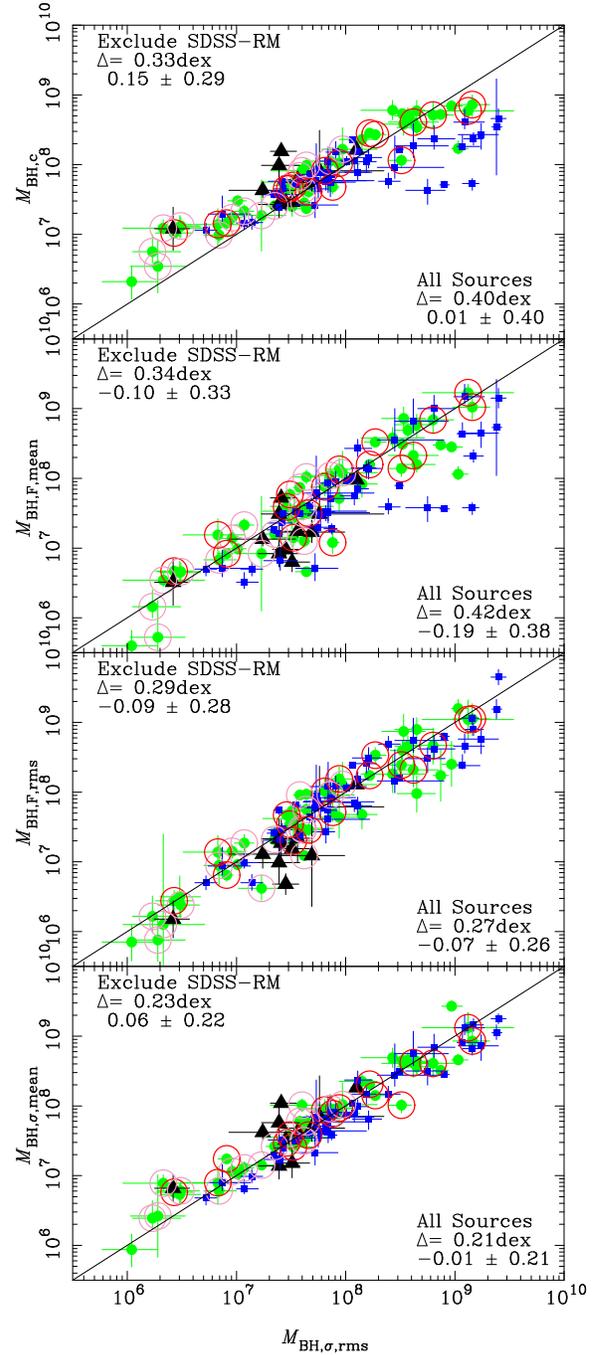}\hfill
\caption{$M_{\rm BH,c}$, $M_{\rm BH,F,mean}$, $M_{\rm BH,F,rms}$ and $M_{\rm BH,\sigma,mean}$ versus $M_{\rm BH,\sigma,rms}$. The green points represent the BentzSample and the black triangles represent the RM AGNs observed by SEAMBH collaboration, the RM AGNs observed by SDSS-RM denote by blue squares. The solid lines are 1:1. The red circles denote RM AGNs with measured \sst for the calibration of $f$. The offset rms (with respect to solid line), the mean value and its standard deviation of the offset for all sources whether excluding SDSS-RM AGNs or not are demonstrated in the top-left/bottom-right corner. }
\label{fig4}
\end{figure}

For our total sample of 120 RM AGNs, we  also use different velocity tracers from the broad \hb line and corresponding $f$ to calculate the SMBH masses.  Fig. \ref{fig4} shows the comparison of the SMBH masses for 106/120 RM AGNs from  different velocity tracers and corresponding $f$ with that from \sr. 
In Table \ref{tab1}, 
Col. (8) is the black hole mass $M_{\rm BH,F,mean}$ derived from the \fwm and corresponding virial factor \ffm = 1.12. 
Col. (9) is  $M_{\rm BH,\sigma,mean}$ derived from the \sm and corresponding virial factor \fsm = 5.50. 
Col. (10) is $M_{\rm BH,F,rms}$ derived from the \fwr and corresponding virial factor \ffm = 1.51 . 
Col. (11) is  $M_{\rm BH,\sigma,rms}$ derived from the \sr and corresponding virial factor \ffm = 6.61 . 
Col. (12) is $M_{\rm BH,c}$ derived from \fwm and corresponding virial factor $\fc \propto \fwm^{-1.10}$. 
It is found that the mass from \sm has the smallest offset rms (with respect to the mass from \sr) of 0.23 dex or 0.21 dex for excluding SDSS-RM sources or not (see the bottom panel in Fig. \ref{fig4}). The red circles denote RM AGNs with measured \sst for the $f$ calibration in Section 3. The rms, mean, and standard deviation of the offset are summarized in Table \ref{tab5}. 

With respect to the mass from \sst or that from the \sr, it is found that we can obtain the SMBH masses from the \sm with the smallest offset rms of 0.38 dex (the fourth panel in Fig. \ref{fig3}) or 0.23 dex (the bottom panel in Fig. \ref{fig4}), respectively, excluding SDSS-RM AGNs. It implies that, rather then the \hb FWHM, the line dispersion \sm from the mean spectrum is preferred to calculate the SMBH mass from the single-epoch spectrum.

For the total sample, using the FWHM-based factor instead of the mean factor, the improvement of mass calculation (from 0.34 dex to 0.32 dex, see top two panels in Fig. \ref{fig4}) are not apparently more significant than that for the subsample of 36 AGNs with \sst (from 0.52 dex to 0.39 dex, see top two panels in Fig. \ref{fig3}). Considering the large coverage of the \hb FWHM in the total sample, especially for SDSS-RM sources, leads to a excessive correction of FWHM-based $f$ (e.g., NLS1s), which leads to a large offset rms for \mbh in top panel in Fig. \ref{fig4}. 

\subsection{The velocity tracer of the BLR clouds}
 \begin{figure*}
 \centering
\includegraphics[angle=-90,width=6.5in]{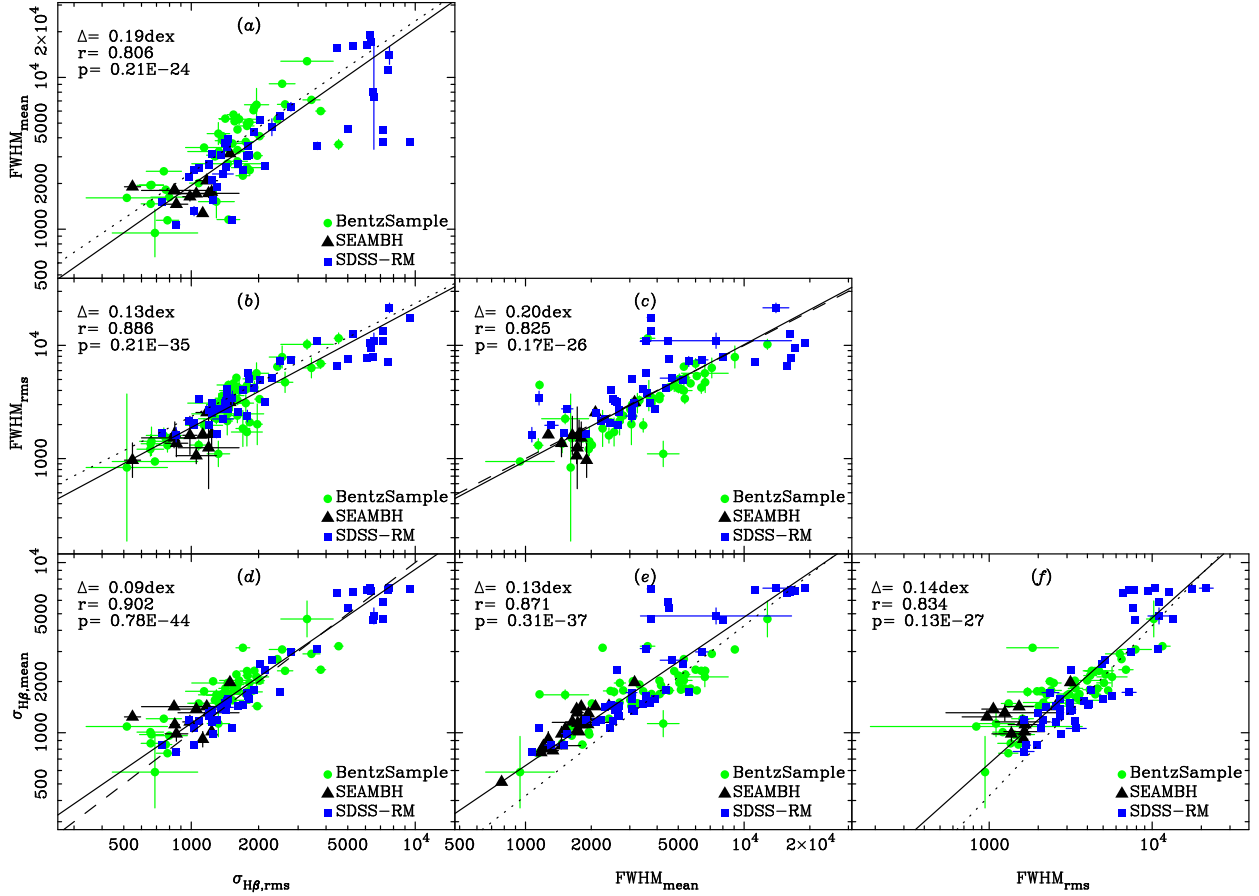}\hfill
\caption{Pairwise correlations among four different velocity tracers. The symbols are the same as Fig. \ref{fig4}. The dashed lines in (c) and (d) are the 1:1. The dotted lines in (a), (b), (e) and (f) correspond to FWHM(H$\beta$)/$\sigma_{\rm H\beta}$ = 2.35 for a Gaussian profile. The solid line in each panel is our best fit derived from BCES bisector. The offset rms respect to the solid line, Spearman correlation coefficient $r_s$ and probability of the null hypothesis $p_{\rm null}$ are shown in the left corner of each panel.
 }
\label{fig5}
\end{figure*}

 \begin{figure}
\includegraphics[angle=-90,width=3.0in]{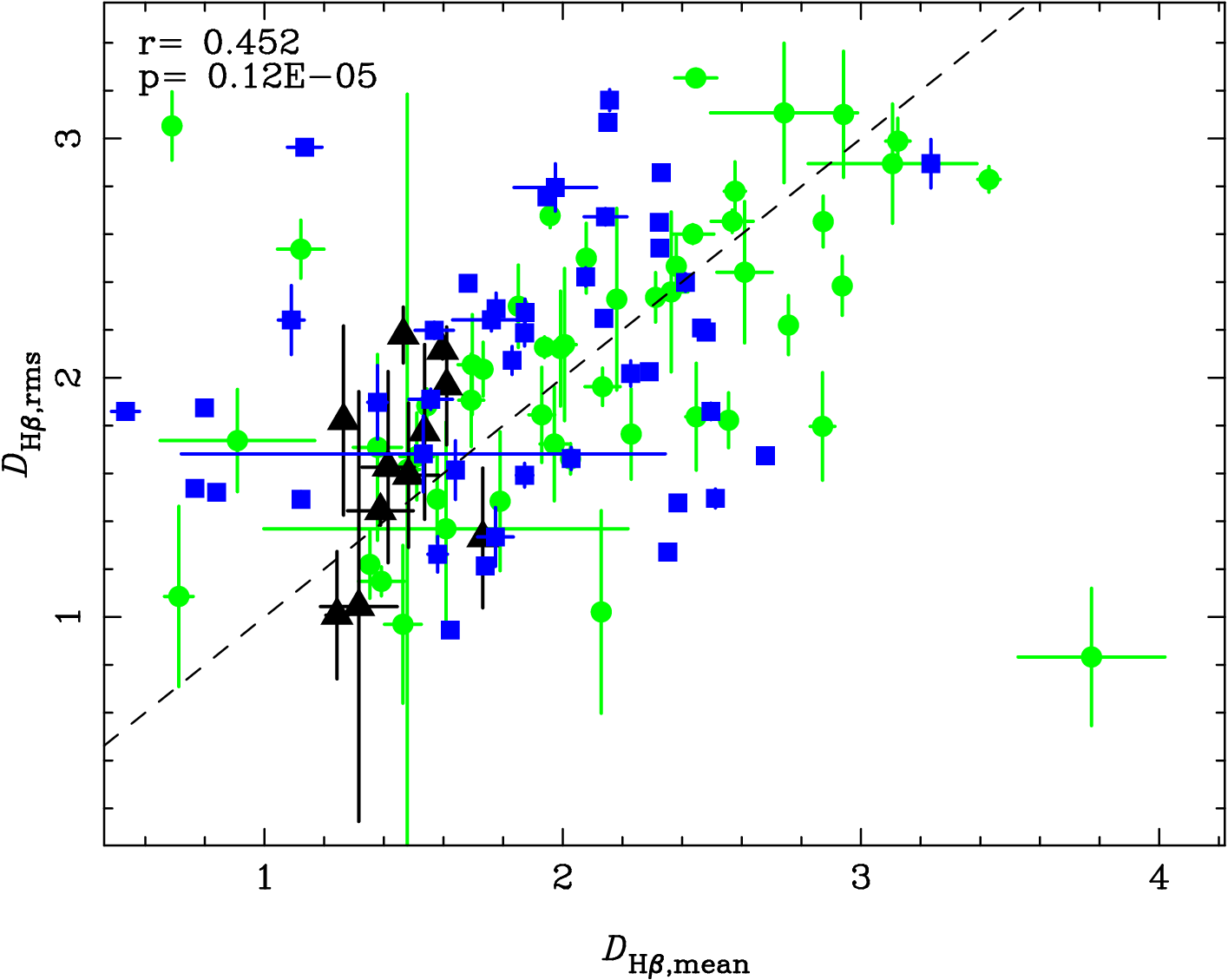}\hfill
\caption{Comparison between \dhbr and \dhbm. The dashed line is the 1:1 and the symbols are the same to Fig. \ref{fig4}.}
\label{fig6}
\end{figure}

The above difference of the SMBH masses is from the different velocity tracer of the BLR clouds $\Delta V$  and the corresponding $f$ in equation (\ref{eq2}). $\Delta V$ can be derived from the broad \hb profile in the mean or rms spectrum. It was found that the scatter in the virial relationship between \sr and the \hb time lag $\tau$ is smallest for several AGNs with multiple mappings, i.e., NGC 5548, NGC 2783, NGC 7469, 3C 390.3 \citep{Pe04}. The line dispersion \sr from the rms spectrum was suggested to be the best tracer \citep{Pe04}. For a Gaussian profile, $\rm FWHM(\hb)/\sigma_{\hb}$ is 2.35. In the logarithm space, the slope for the relation between FWHM(\hb) and $\sigma_{\rm \hb}$ is one. For our total compiles sample, \fwm ranges from 778 \kms to 17112 \kms. The subsample of SEAMBHs has small \fwm locating the bottom-left corner of panel (a) in Fig. \ref{fig5} and the subsample of SDSS-RM tends to have large \fwm. 

In this subsection, we compare the correlations among these four kinds of BLRs velocity tracers and calculate the scatter using other velocity tracers instead of \sr.  Fig. \ref{fig5} shows the pairwise correlations among four different velocity tracers (i.e., \fwm,\fwr,\sm and \sr). The dashed lines in (c) and (d) are the 1:1. The dotted lines in (a), (b), (e) and (f) correspond FWHM(H$\beta$)/$\sigma_{\rm H\beta}$ = 2.35 for a Gaussian profile. We adopt BCES bisector \citep{AB96} to do the linear fit for these relations. The solid line in each panel is our best fit. All the correlations are significant with $p_{\rm null} < 10^{-24}$. The fitting results are summarized in Table \ref{tab4}. The slopes in panels (a) and (b) are larger than 1.0 and slopes in panels (e) and (f) are less than 1.0. 
For NLS1s, FWHM tends to be less than 2.35 $\times$ \shb (see panels (a), (b), (e) and (f)) \citep[e.g.,][]{Co06, bian2008, Yu20}.
The panels (a) and (c) in Fig. \ref{fig5}, i.e., \fwm vs. \sr and \fwm vs. \fwr,  show larger scatter than others (0.19 dex vs. 0.20 dex). The offset rms of \sm respect to \sr is the smallest (0.09 dex). We can use the relation between \sm and \sr to estimate the \sr from \sm (in a single-epoch spectrum) as the velocity of the BLR clouds, with an additional scatter of 0.18 dex for the single-epoch \mbh (in equation \ref{eq2}).

In Fig. \ref{fig6}, we also show the relation of \hb profile ($D_{\rm \hb}=\rm FWHM(\hb)/\sigma_{\hb}$) between the mean and the rms spectrum \citep[e.g.,][]{Co06, Yu20}. There is medium-strong correlation between them with $r_s=0.452, p_{\rm null}=0.12\times 10^{-5}$. The larger scatter suggests the difference of the \hb profile between the mean and rms spectra.

\section{Conclusions}
For a compiled sample of 120 RM AGNs, their SMBH masses are calculated using different kinds of velocity tracers of the BLR clouds from the broad \hb profile in the mean or rms spectrum  and the corresponding virial factor $f$, as well as through the \ms relation for its subsample of 36 RM AGNs with measured \sst. The main conclusions are summarized as follows:

\begin{itemize}
\item For 36 RM AGNs with measured \sst, considering the intrinsic scatter in the \ms relation for AGNs, the multivariate linear regression technique is used to calibrate the mean value of the factor $f$ for different tracers of cloud velocity from the \hb line profile measured from the mean or rms spectrum, as well as the FWHM-based factor. The calibrations of the factor $f$ for 17 low-z RM AGNs with E/CBs are $\ffm = 1.12 \pm 0.17$, $\ffr = 1.51 \pm 0.20$, $\fsm = 5.50 \pm 0.74$ and $\fsr = 6.61 \pm 0.81$, which is consistent well with that by \cite{HK14}. We also confirm the dependence of variable $f$ on the FWHM(\hb), i.e., $\ffm \propto \fwm^{-1.10\pm 0.40}$ \citep{Yu19}. Including high-z SDSS-RM AGNs and/or PBs, we also find similar results. The FWHM-based variable $f$ suggests the non-virial \mbh calculated from the \hb FWHM.


\item  For 106/120 RM AGNs, the SMBH masses from different kinds of velocity tracers and the corresponding virial factors are compared with that from \sst or that from the assumed best tracer \sr. With respect to the mass from \sst or that from the \sr, it is found that we can obtain the SMBH masses from \sm with the smallest offset rms of 0.38 dex or 0.23 dex, respectively, excluding SDSS-RM AGNs. It implies that, with respect to the \hb FWHM, we prefer the line dispersion \sm from the mean spectrum to calculate the single-epoch SMBH mass.

\item Using the FWHM-based factor, we can improve \mbh calculation from the \hb FWHM and the mean factor, with a decreased offset rms from 0.52 dex to 0.39 dex with respect to \mbh from \sst for the calibration subsample of 36 AGNs with \sst. The scatter of 0.39 dex is almost the same as that derived from \sm and the mean factor. For the total sample, considering the FWHM-based factor instead of the mean factor, the improvement of mass calculation are not apparently more significant than that for the subsample of 36 AGNs with \sst, which is due to the excessive correction of $f$ for AGNs with large or small FWHM.

\item For the total sample of RM AGNs, we compare the assumed best velocity tracer of \sr from the rms spectrum with other tracer of \fwm or \sm from the mean spectrum. Their relations are obtained and the velocity of \fwm or \sm from the mean spectrum can be scaled to \sr.  It would be used in the calculation of the SMBH mass from a single-epoch spectrum. Using \sm instead of \sr, we find that an additional scatter of 0.18 dex in \mbh is introduced.


\end{itemize}

\acknowledgments

We are very grateful to R. F. Green for his instructive comments.
We are also very grateful to the anonymous referee for her/his instructive comments which significantly improved the content of the paper. 
This work is supported by the National Key Research and Development Program of China (No. 2017YFA0402703). This work has been supported by the National Science Foundations of China (Nos. 11973029 and 11873032).

\newpage
\appendix
\renewcommand\thefigure{\Alph{section}\arabic{figure}}   

\section{The calibration of the virial factor for RM AGNs including SDSS-RM AGNs}
\setcounter{figure}{0}  
\begin{figure*}
\includegraphics[angle=-90,width=3.4in]{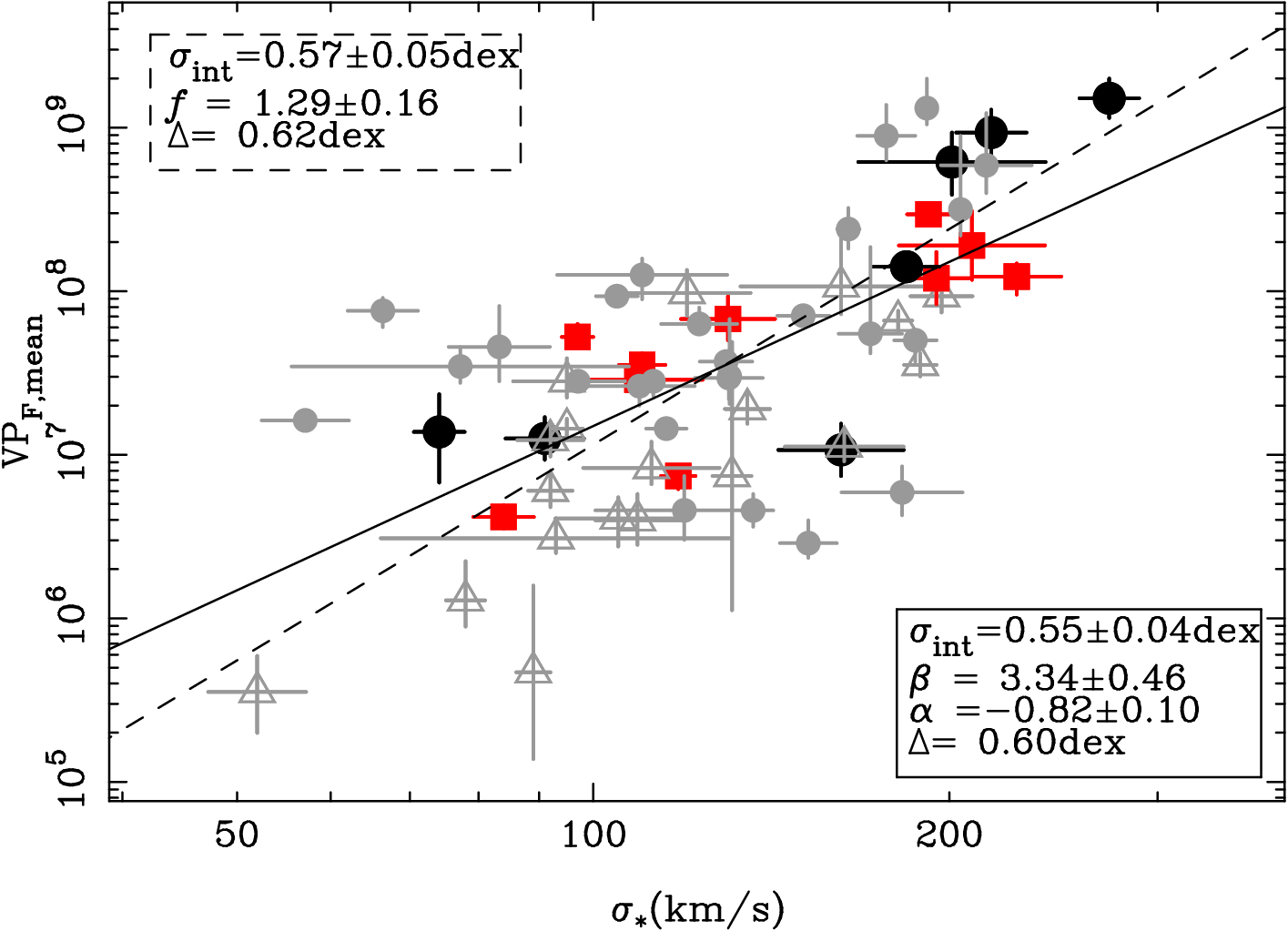}\hfill
\includegraphics[angle=-90,width=3.4in]{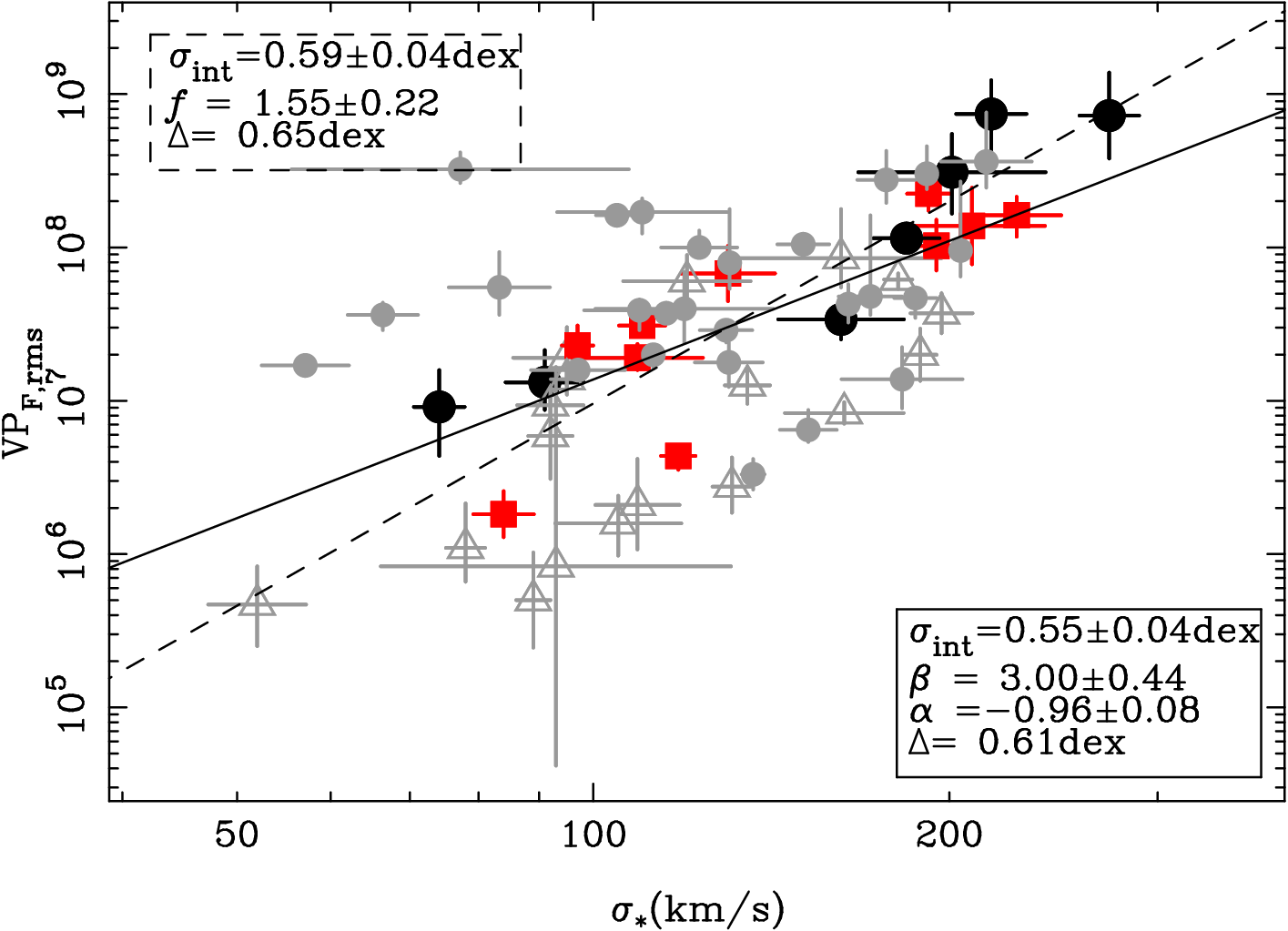}\vfill
\includegraphics[angle=-90,width=3.4in]{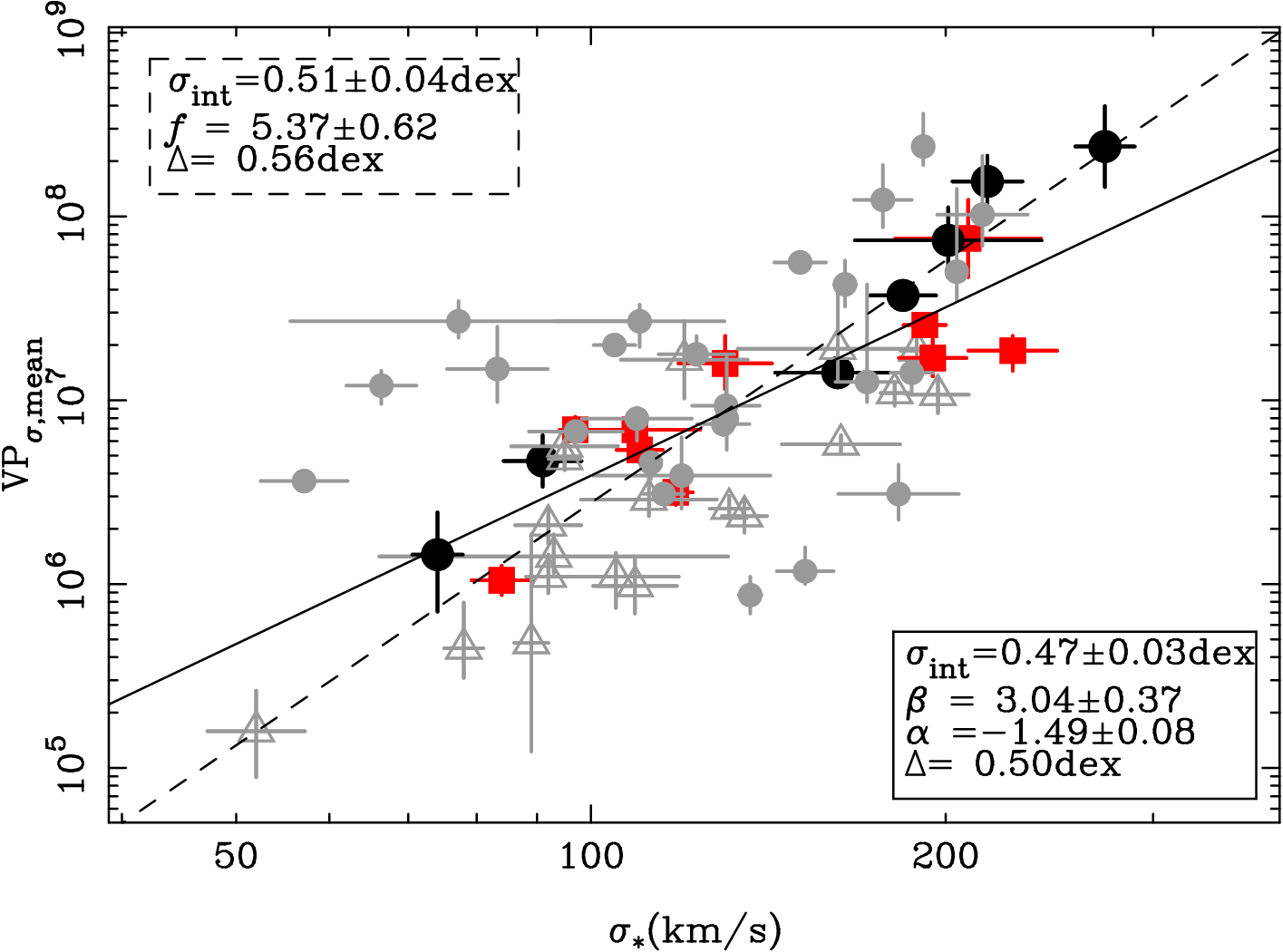}\hfill
\includegraphics[angle=-90,width=3.4in]{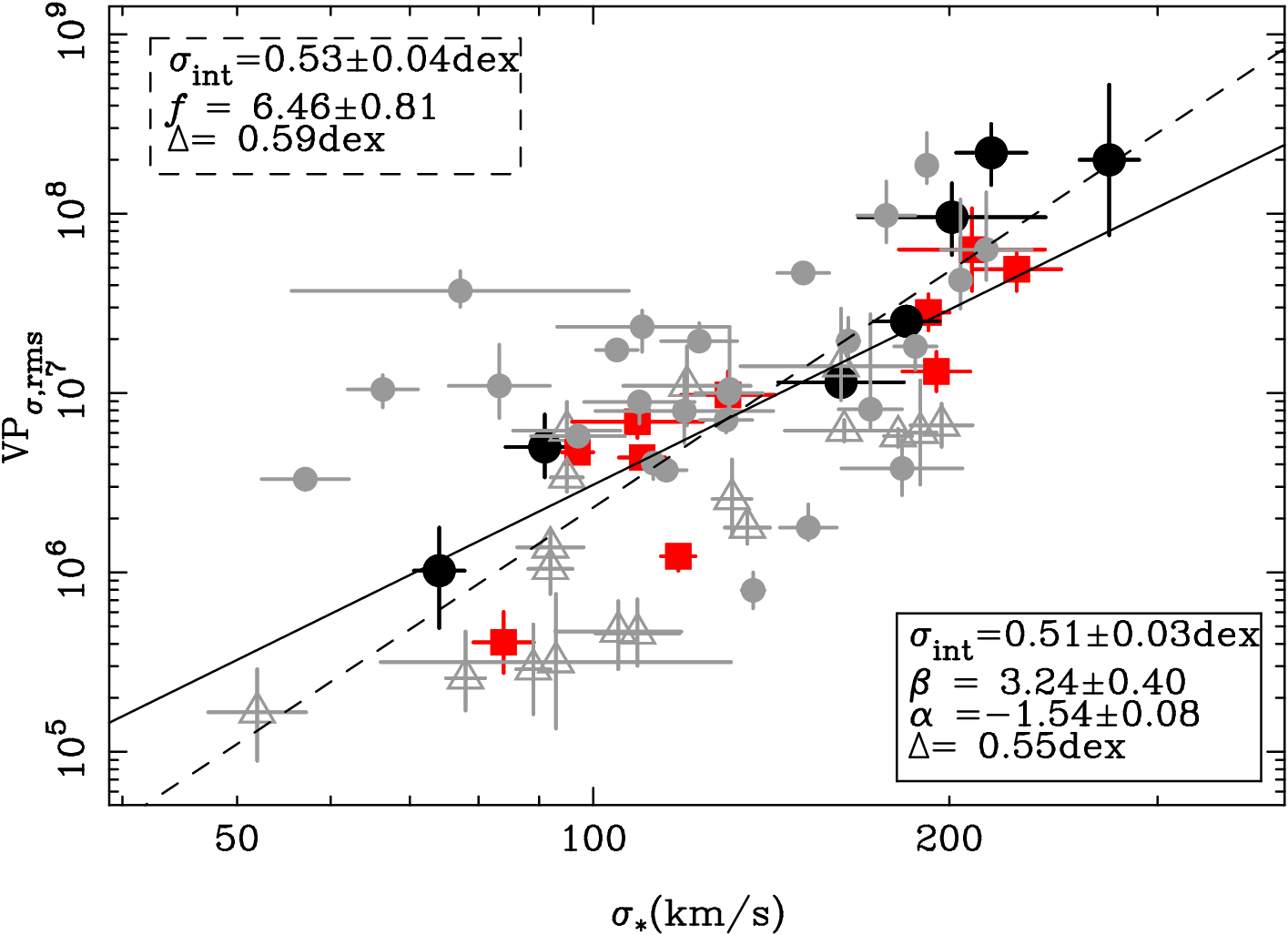}\hfill

\caption{Four kinds of VPs versus \sst including SDSS-RM AGNs. Top left: \vpfm versus \sst. Top right: \vpsm versus \sst. Bottom left: \vpfr versus \sst. Bottom right: \vpsr versus \sst. The symbols are the same as Fig. \ref{fig1}, as well as the contents in the insert boxes. SDSS-RM AGNs are shown as gray circles. Our best-fit line for all AGNs is shown as the black solid line. The detail about the best-fit line is shown in the bottom right corner. Fixing the slope ($\beta$ = 4.38) same as the \ms relation for quiescent galaxies \citep{KH13}, the dashed line is our best fit. The detail is presented in the upper left corner.} 
\label{fig1-1}
\end{figure*}

\begin{figure*}
\includegraphics[angle=-90,width=3.5in]{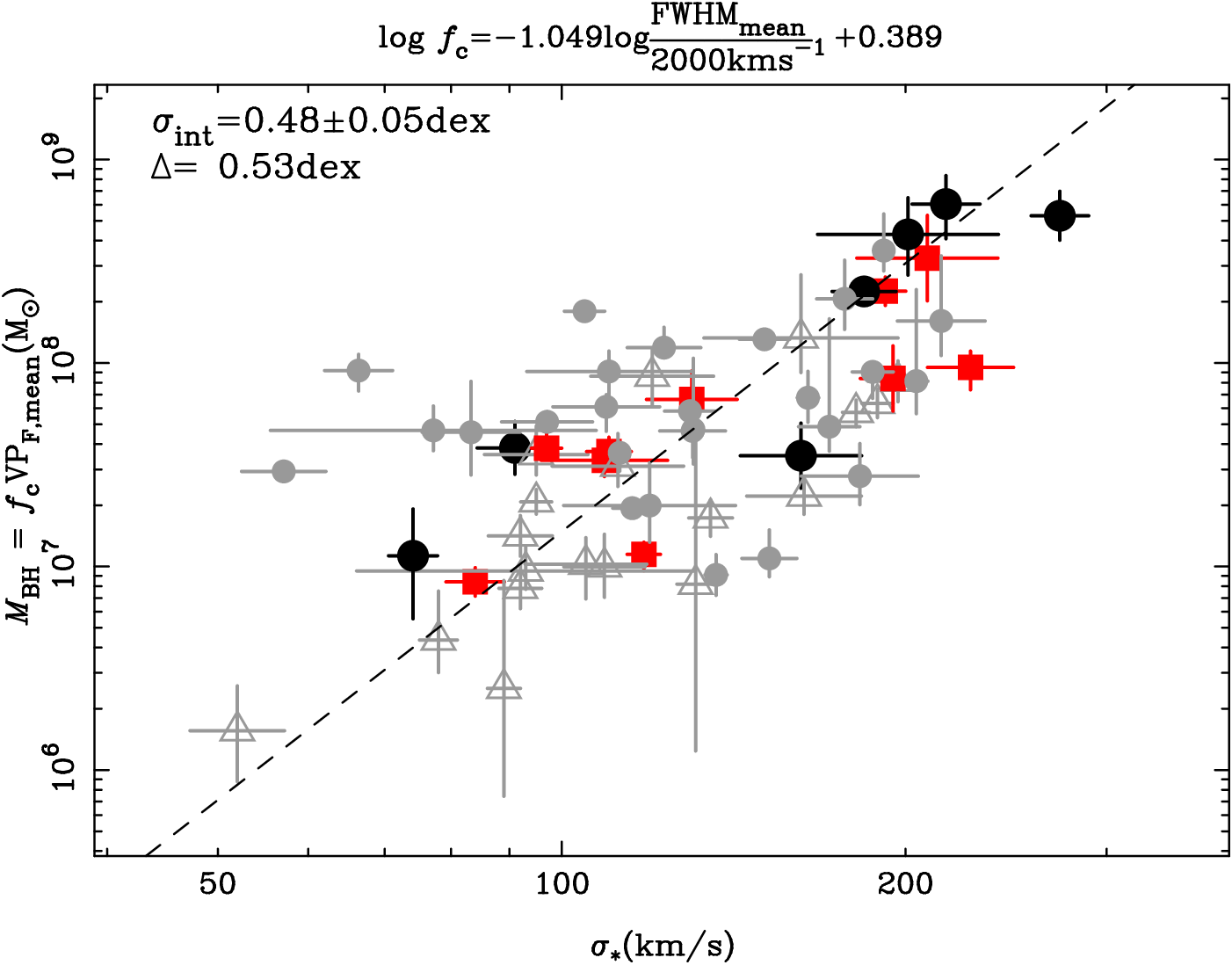}
\includegraphics[angle=-90,width=3.5in]{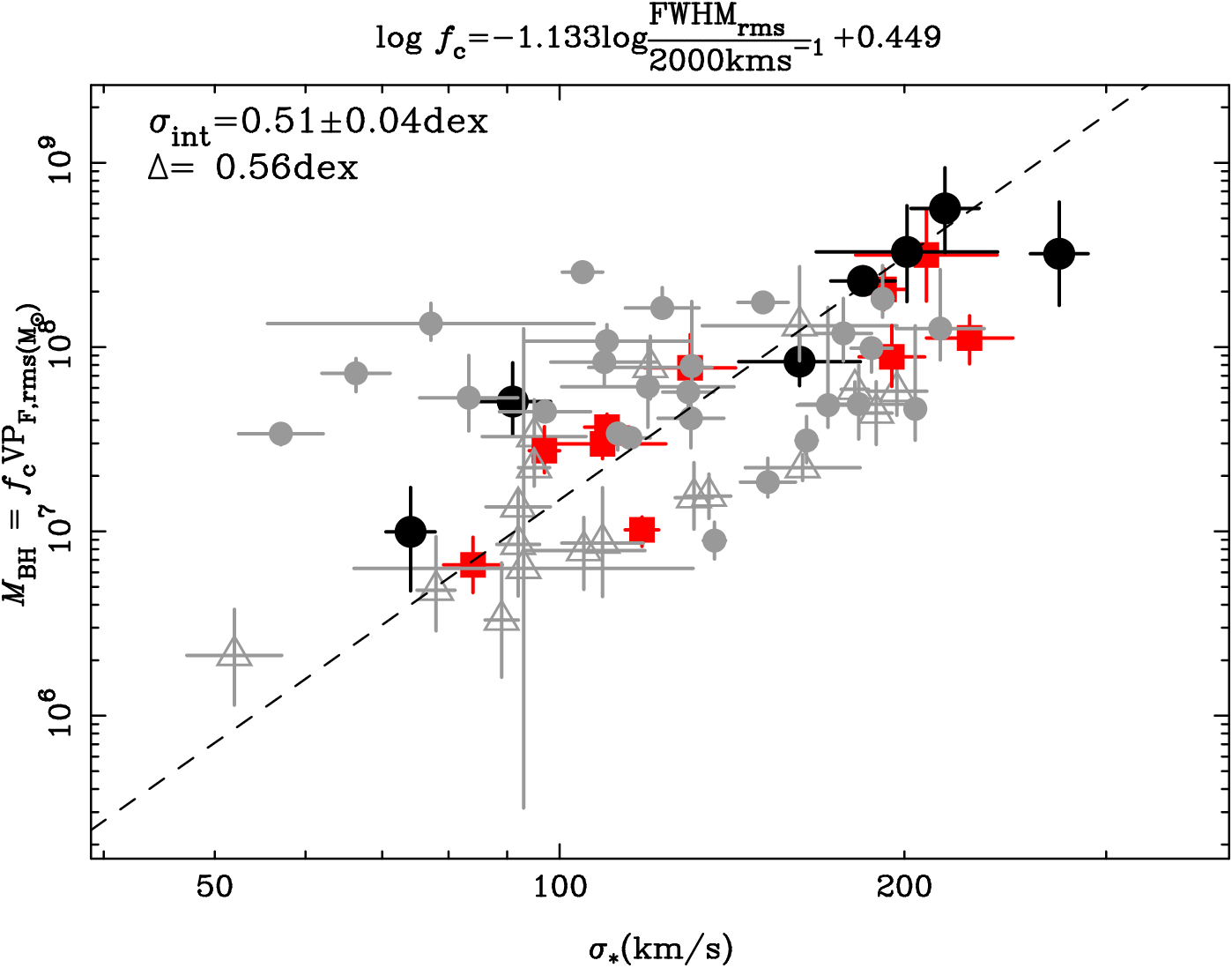}\hfill
\caption{The \ms relation including SDSS-RM AGNs. Left: \mbh are calculate from \vpfm, and we show the virial factor corrected by \fwm on the top of the figure. Right: \mbh are calculate from \vpfr, and we also show the virial factor corrected by \fwr on the top of the figure. The symbols are the same as Fig. \ref{fig1-1}.}
\label{fig2-1}
\end{figure*}

In Section 3, we use the subsample of 17 E/CB out of 36 low-z RM-AGN with measured \sst to calibrate the varial factor $f$. \cite{Sh15} showed no evidence of \ms evolution up to redshift 1 for SDSS-RM AGNs. \cite{Li2020} also suggested no strong evidence of evolution in the $M_{\rm BH}- M_{\star,bulge}$ relation to $z \sim$ 0.6. In this Appendix, including 26 SDSS-RM AGNs with measured \sst, we also calibrate the virial factor $f$ as we do in Section 3. The results are shown in Table \ref{tab3}. For the entire calibration sample of 62 RM-AGNs with \sst including these SDSS-RM AGNs, we calibrate the mean $f$ based on different \hb velocity tracers. In Fig. \ref{fig1-1}, for a fixed slope of $\beta=4.38$ (dashed lines), it is found that the intrinsic scatters $\sigma_{\rm int}$ are lager than the results shown in Fig. \ref{fig1} excluding SDSS-RM AGNs and PBs, i.e., 0.19, 0.22, 0.3, 0.24 dex for \fwm, \sm, \fwr, \sr, respectively. It is almost double the $\sigma_{\rm int}$ founded in Fig. \ref{fig1} excluding SDSS-RM AGNs. However, considering the error, the values of $f$ are almost the same to that shown in Fig. \ref{fig1} excluding SDSS-RM AGNs. The rms offset is also larger.   
Releasing the fixed slope, we give the results in the right corner in reach panels in Fig. \ref{fig1-1} (see the solid lines). $\sigma_{\rm int}$ and $\Delta$ are larger than for excluding SDSS-RM AGNs. The slope is flatter than for excluding SDSS-RM AGNs. 

Including SDSS-RM AGNs, we also give the results on variable $f$ instead of the constant $f$ alike in subsection 3.3 (Fig. \ref{fig2-1}). Including SDSS-RM AGNs, the \fwm-based $f$ is almost the same with a slightly larger $\sigma_{\rm int}$. However, the \fwr-based $f$ has a large difference, i.e., the slope from -0.5 to -1.13. 

For 43 RM-AGNs with \sst excluding PBs (we treat 26 SDSS-RM AGNs as classical bulges), we also do the calibration of $f$ like above (see Table \ref{tab3}). For a fixed slope of $\beta=4.38$, the mean values of $f$ become smaller, from $1.29\pm 0.16, 1.55\pm 0.22, 5.37\pm 0.62, 6.46\pm 0.81$ to $1.00\pm 0.16, 1.10\pm 0.17, 4.17\pm 0.58, 4.68\pm 0.60$ for \fwm, \sm, \fwr, \sr, respectively. Releasing the fixed slope, the fitting slopes become flatter, from $3.34\pm 0.46, 3.00\pm 0.44, 3.04\pm 0.37, 3.24\pm 0.40$ to $2.69\pm 0.41, 2.32\pm 0.50, 2.50\pm 0.38, 2.67\pm 0.45 $ for \fwm, \sm, \fwr, \sr, respectively. For FWHM-based $f$, the slope in the relation between $f$ and FWHM changes slightly within the uncertainty of $1~ \sigma$.

Considering the almost same mean $f$, the same $f$ relation with \fwm, and the larger scatter including SDSS-RM AGNs, we prefer the results for 17 RM-AGNs excluding SDSS-RM AGNs and PBs shown in Section 3 (see Table \ref{tab3}).

\newpage



\newpage
\begin{table*}
\caption{The properties of 120 RM AGNs. }
\centering
\label{tab1}
\begin{lrbox}{\tablebox}
\begin{tabular}{llllllllllllll}
\hline
Name&$\tau$&$\log \lv$&\fwm&\sm&\fwr&\sr&$\log M_{\rm BH,F,mean}$&$\log M_{\rm BH,\sigma,mean}$&$\log M_{\rm BH,F,rms}$&$\log M_{\rm BH,\sigma,rms}$&$\log M_{\rm BH,c}$&Ref.\\
         &(days)&$\log \ergs$&(\kms)&(\kms)&(\kms)&(\kms)&($\log \msun$)&($\log \msun$)&($\log \msun$)&($\log \msun$)&($\log \msun$)\\
 (1)&(2)&(3)&(4)&(5)&(6)&(7)&(8)&(9)&(10)&(11)&(12)&(13)\\
\hline
\multicolumn{13}{c}{SEAMBH}\\
Mrk335&                  $  8.7_{-  1.9}^{+  1.6}$&$ 43.69\pm  0.06$&$ 2096\pm 170$&$ 1470\pm  50$&$\cdots$&      $\cdots$&      $ 6.92_{- 0.18}^{+ 0.16}$&$ 7.30_{- 0.15}^{+ 0.13}$&$\cdots$&                 $\cdots$&                 $ 7.33_{- 0.18}^{+ 0.16}$&1, 2, 3   \\
Mrk335&                  $ 16.8_{-  4.2}^{+  4.8}$&$ 43.76\pm  0.06$&$ 1792\pm   3$&$ 1380\pm   6$&$ 1629\pm 145$&$  917\pm  52$&$ 7.07_{- 0.16}^{+ 0.19}$&$ 7.54_{- 0.16}^{+ 0.19}$&$ 7.12_{- 0.20}^{+ 0.22}$&$ 7.26_{- 0.18}^{+ 0.20}$&$ 7.55_{- 0.16}^{+ 0.19}$&2, 3, 4   \\
Mrk335&                  $ 12.5_{-  5.5}^{+  6.6}$&$ 43.84\pm  0.06$&$ 1679\pm   2$&$ 1371\pm   8$&$ 1375\pm 357$&$  948\pm 113$&$ 6.89_{- 0.29}^{+ 0.35}$&$ 7.40_{- 0.29}^{+ 0.35}$&$ 6.84_{- 0.45}^{+ 0.49}$&$ 7.16_{- 0.33}^{+ 0.38}$&$ 7.40_{- 0.29}^{+ 0.35}$&2, 3, 4   \\
Mrk335&                  $ 14.3_{-  0.7}^{+  0.7}$&$ 43.74\pm  0.06$&$ 1724\pm 236$&$ 1542\pm  66$&$ 1025\pm  35$&$ 1293\pm  64$&$ 6.97_{- 0.18}^{+ 0.18}$&$ 7.56_{- 0.06}^{+ 0.06}$&$ 6.65_{- 0.06}^{+ 0.06}$&$ 7.49_{- 0.07}^{+ 0.07}$&$ 7.47_{- 0.18}^{+ 0.18}$&1, 2, 5   \\
\hline
\end{tabular}
\end{lrbox}
\scalebox{0.6}{\usebox{\tablebox}}
\end{table*}
\begin{table*}
\centering
\begin{lrbox}{\tablebox}
\begin{tabular}{llllllllllllll}
\hline
Name&$\tau$&$\log \lv$&\fwm&\sm&\fwr&\sr&$\log M_{\rm BH,F,mean}$&$\log M_{\rm BH,\sigma,mean}$&$\log M_{\rm BH,F,rms}$&$\log M_{\rm BH,\sigma,rms}$&$\log M_{\rm BH,c}$&Ref.\\
         &(days)&$\log \ergs$&(\kms)&(\kms)&(\kms)&(\kms)&($\log \msun$)&($\log \msun$)&($\log \msun$)&($\log \msun$)&($\log \msun$)\\
 (1)&(2)&(3)&(4)&(5)&(6)&(7)&(8)&(9)&(10)&(11)&(12)&(13)\\
\hline
RMID779&                 $ 11.8_{-  1.5}^{+  0.7}$&$ 43.12\pm  0.00$&$ 2670\pm  17$&$ 1249\pm   4$&$ 2709\pm  55$&$ 1205\pm   9$&$ 7.26_{- 0.08}^{+ 0.04}$&$ 7.30_{- 0.08}^{+ 0.04}$&$ 7.41_{- 0.09}^{+ 0.05}$&$ 7.34_{- 0.08}^{+ 0.04}$&$ 7.56_{- 0.08}^{+ 0.04}$&9, 10, 11 \\
RMID781&                 $ 75.2_{-  3.3}^{+  3.2}$&$ 43.64\pm  0.02$&$ 2515\pm  26$&$ 1169\pm   5$&$ 3340\pm  82$&$ 1089\pm  22$&$ 8.02_{- 0.03}^{+ 0.03}$&$ 8.04_{- 0.03}^{+ 0.03}$&$ 8.39_{- 0.04}^{+ 0.04}$&$ 8.06_{- 0.04}^{+ 0.04}$&$ 8.34_{- 0.03}^{+ 0.03}$&9, 10, 11 \\
RMID782&                 $ 20.0_{-  3.0}^{+  1.1}$&$ 43.94\pm  0.00$&$ 3070\pm  49$&$ 1378\pm   6$&$ 2730\pm 137$&$ 1353\pm  23$&$ 7.61_{- 0.10}^{+ 0.04}$&$ 7.61_{- 0.10}^{+ 0.04}$&$ 7.64_{- 0.12}^{+ 0.08}$&$ 7.67_{- 0.10}^{+ 0.04}$&$ 7.84_{- 0.10}^{+ 0.04}$&9, 10, 11 \\
RMID790&                 $  5.5_{-  2.1}^{+  5.7}$&$ 43.29\pm  0.00$&$17112\pm  81$&$ 6813\pm  13$&$ 9448\pm 367$&$ 6318\pm  38$&$ 8.55_{- 0.25}^{+ 0.68}$&$ 8.44_{- 0.25}^{+ 0.68}$&$ 8.16_{- 0.26}^{+ 0.68}$&$ 8.45_{- 0.25}^{+ 0.68}$&$ 7.95_{- 0.25}^{+ 0.68}$&9, 10, 11 \\
RMID840&                 $  5.0_{-  1.4}^{+  1.5}$&$ 43.21\pm  0.00$&$15735\pm  93$&$ 6596\pm  22$&$ 6580\pm  48$&$ 4457\pm  60$&$ 8.43_{- 0.18}^{+ 0.20}$&$ 8.37_{- 0.18}^{+ 0.20}$&$ 7.80_{- 0.18}^{+ 0.20}$&$ 8.11_{- 0.18}^{+ 0.20}$&$ 7.88_{- 0.18}^{+ 0.20}$&9, 10, 11 \\
\hline
\end{tabular}
\end{lrbox}
\scalebox{0.6}{\usebox{\tablebox}}
\end{table*}
\begin{table*}
\centering
\begin{lrbox}{\tablebox}
\begin{tabular}{llllllllllllll}
\hline
Name&$\tau$&$\log \lv$&\fwm&\sm&\fwr&\sr&$\log M_{\rm BH,F,mean}$&$\log M_{\rm BH,\sigma,mean}$&$\log M_{\rm BH,F,rms}$&$\log M_{\rm BH,\sigma,rms}$&$\log M_{\rm BH,c}$&Ref.\\
         &(days)&$\log \ergs$&(\kms)&(\kms)&(\kms)&(\kms)&($\log \msun$)&($\log \msun$)&($\log \msun$)&($\log \msun$)&($\log \msun$)\\
 (1)&(2)&(3)&(4)&(5)&(6)&(7)&(8)&(9)&(10)&(11)&(12)&(13)\\
\hline
Mrk1511&                 $  5.7_{-  0.8}^{+  0.9}$&$ 43.16\pm  0.06$&$ 4171\pm 137$&$ 1899\pm  95$&$\cdots$&      $\cdots$&      $ 7.34_{- 0.10}^{+ 0.11}$&$ 7.34_{- 0.11}^{+ 0.12}$&$\cdots$&                 $\cdots$&                 $ 7.42_{- 0.10}^{+ 0.11}$&1, 2, 22  \\
Mrk290&                  $  8.7_{-  1.0}^{+  1.2}$&$ 43.17\pm  0.06$&$ 4543\pm 227$&$ 1769\pm  88$&$ 4270\pm 157$&$ 1609\pm  47$&$ 7.59_{- 0.10}^{+ 0.11}$&$ 7.47_{- 0.10}^{+ 0.11}$&$ 7.67_{- 0.09}^{+ 0.10}$&$ 7.46_{- 0.09}^{+ 0.10}$&$ 7.63_{- 0.10}^{+ 0.11}$&1, 2, 4, 1\\
PG1613+658&              $ 40.1_{- 15.2}^{+ 15.0}$&$ 44.77\pm  0.02$&$ 9074\pm 103$&$ 3085\pm  33$&$ 7897\pm1792$&$ 2547\pm 342$&$ 8.86_{- 0.25}^{+ 0.25}$&$ 8.61_{- 0.25}^{+ 0.25}$&$ 8.87_{- 0.39}^{+ 0.39}$&$ 8.53_{- 0.30}^{+ 0.30}$&$ 8.57_{- 0.25}^{+ 0.25}$&1, 2, 3, 4\\
PG1617+175&              $ 71.5_{- 33.7}^{+ 29.6}$&$ 44.39\pm  0.02$&$ 6641\pm 190$&$ 2313\pm  69$&$ 4718\pm 991$&$ 2626\pm 211$&$ 8.84_{- 0.31}^{+ 0.27}$&$ 8.61_{- 0.31}^{+ 0.27}$&$ 8.67_{- 0.41}^{+ 0.39}$&$ 8.80_{- 0.33}^{+ 0.29}$&$ 8.70_{- 0.31}^{+ 0.27}$&1, 2, 3, 4\\
PG1700+518&              $251.8_{- 38.8}^{+ 45.9}$&$ 45.59\pm  0.01$&$ 2252\pm  85$&$ 3160\pm  93$&$ 1846\pm 682$&$ 1700\pm 123$&$ 8.45_{- 0.11}^{+ 0.13}$&$ 9.43_{- 0.11}^{+ 0.13}$&$ 8.40_{- 0.49}^{+ 0.50}$&$ 8.97_{- 0.14}^{+ 0.15}$&$ 8.82_{- 0.11}^{+ 0.13}$&1, 2, 3, 4\\
\hline
\end{tabular}
\end{lrbox}
\scalebox{0.6}{\usebox{\tablebox}}
\\
Note. For a object with multiple measurements, we highlighted the name in boldface. \\
Reference: 1: \cite{Du15}, 2: \cite{Du16a}, 3: \cite{Co06}, 4: \cite{Be13}, 5: \cite{Gr12}, 6: \cite{Be09a},  7: \cite{Du18}, 8: \cite{Du16b},  9: \cite{Sh15},  10: \cite{Gr17},  11: \cite{Shen19},   12:  \cite{Be09b},  13: \cite{Denny2010},   14: \cite{Pe00},   15: \cite{Be06}, 16: \cite{Fa17} 17: \cite{Zhang19}  18: \cite{Denny2006}   19: \cite{Be14},   20: \cite{Lu16},  21: \cite{Pei17},  22: \cite{Ba13} ,  23: \cite{Pei14},  24: \cite{Pe14}, 25:\cite{HK14},   26: \cite{Ba15},   27: \cite{Williams2018}, 28: \cite{Be16a},  29: \cite{Be16b}\\
(This table is available in its entirety in machine-readable form.)
\end{table*}

\begin{table*}
\caption{The properties of 36 low-z RM AGNs and 26 SDSS-RM AGNs for calibrating $f$ }
\centering
\label{tab2}
\begin{lrbox}{\tablebox}
\begin{tabular}{llllllllllllll}
\hline
Name&Bulge&\sst&$\tau$&$\log \lv$&$\log \vpfm$&$\log \vpsm$&$\log \vpfr$&$\log \vpsr$&Ref.\\
         &Type&(\kms)&(days)&$\log \ergs$&($\log \msun$)&($\log \msun$)&($\log \msun$)&($\log \msun$)&\\
 (1)&(2)&(3)&(4)&(5)&(6)&(7)&(8)&(9)&(10)\\
\hline
Fairall9&                CB&$228  \pm20  $&$ 17.4_{-  4.3}^{+  3.2}$&$ 43.98\pm  0.04$&$ 8.09_{- 0.11}^{+ 0.08}$&$ 7.27_{- 0.11}^{+ 0.08}$&$ 8.21_{- 0.14}^{+ 0.12}$&$ 7.69_{- 0.12}^{+ 0.09}$&1,2,3,4    \\
Mrk590&                  &  &              $ 20.7_{-  2.7}^{+  3.5}$&$ 43.59\pm  0.06$&$ 7.50_{- 0.06}^{+ 0.07}$&$ 7.18_{- 0.06}^{+ 0.07}$&$ 7.05_{- 0.31}^{+ 0.31}$&$ 6.40_{- 0.10}^{+ 0.11}$&2,3,4      \\
Mrk590&                  &  &              $ 14.0_{-  8.8}^{+  8.5}$&$ 43.14\pm  0.09$&$ 7.58_{- 0.29}^{+ 0.28}$&$ 7.11_{- 0.27}^{+ 0.26}$&$ 7.25_{- 0.27}^{+ 0.26}$&$ 7.01_{- 0.27}^{+ 0.26}$&2,3,4      \\
Mrk590&                  &  &              $ 29.2_{-  5.0}^{+  4.9}$&$ 43.38\pm  0.07$&$ 7.63_{- 0.08}^{+ 0.08}$&$ 7.34_{- 0.07}^{+ 0.07}$&$ 7.41_{- 0.25}^{+ 0.24}$&$ 6.95_{- 0.09}^{+ 0.09}$&2,3,4      \\
Mrk590&                  &  &              $ 28.8_{-  4.2}^{+  3.6}$&$ 43.65\pm  0.06$&$ 7.55_{- 0.06}^{+ 0.06}$&$ 7.30_{- 0.06}^{+ 0.05}$&$ 7.34_{- 0.18}^{+ 0.18}$&$ 6.91_{- 0.11}^{+ 0.11}$&2,3,4      \\
\bfseries Mrk590&        PB&$189  \pm6   $&$ 25.6_{-  5.7}^{+  5.8}$&$ 43.50\pm  0.21$&$ 7.55_{- 0.07}^{+ 0.07}$&$ 7.27_{- 0.09}^{+ 0.09}$&$ 7.30_{- 0.17}^{+ 0.17}$&$ 6.78_{- 0.29}^{+ 0.29}$&1          \\
3C120&                   &  &              $ 38.1_{- 15.3}^{+ 21.3}$&$ 44.07\pm  0.05$&$ 7.60_{- 0.17}^{+ 0.24}$&$ 7.06_{- 0.17}^{+ 0.24}$&$ 7.56_{- 0.19}^{+ 0.25}$&$ 7.00_{- 0.18}^{+ 0.24}$&2,3,4      \\
3C120&                   &  &              $ 25.9_{-  2.3}^{+  2.3}$&$ 43.94\pm  0.05$&$ 7.01_{- 0.04}^{+ 0.04}$&$ 7.16_{- 0.04}^{+ 0.04}$&$ 7.51_{- 0.16}^{+ 0.16}$&$ 7.06_{- 0.05}^{+ 0.05}$&4,5        \\
\bfseries 3C120&         E& $162  \pm20  $&$ 26.2_{-  3.5}^{+  3.5}$&$ 44.01\pm  0.10$&$ 7.03_{- 0.16}^{+ 0.16}$&$ 7.15_{- 0.04}^{+ 0.04}$&$ 7.53_{- 0.13}^{+ 0.14}$&$ 7.06_{- 0.05}^{+ 0.06}$&1          \\
Ark120&                  &  &              $ 47.1_{- 12.4}^{+  8.3}$&$ 43.98\pm  0.06$&$ 8.53_{- 0.11}^{+ 0.08}$&$ 7.45_{- 0.11}^{+ 0.08}$&$ 8.45_{- 0.12}^{+ 0.09}$&$ 7.55_{- 0.12}^{+ 0.09}$&2,3,4      \\
Ark120&                  &  &              $ 37.1_{-  5.4}^{+  4.8}$&$ 43.63\pm  0.08$&$ 8.45_{- 0.06}^{+ 0.06}$&$ 7.40_{- 0.06}^{+ 0.06}$&$ 8.31_{- 0.07}^{+ 0.06}$&$ 7.41_{- 0.07}^{+ 0.06}$&2,3,4      \\
\bfseries Ark120&        CB&$192  \pm8   $&$ 39.7_{-  7.9}^{+  7.4}$&$ 43.85\pm  0.24$&$ 8.47_{- 0.07}^{+ 0.07}$&$ 7.41_{- 0.06}^{+ 0.06}$&$ 8.35_{- 0.11}^{+ 0.11}$&$ 7.45_{- 0.10}^{+ 0.10}$&1          \\
Mrk79&                   &  &              $  9.0_{-  7.8}^{+  8.3}$&$ 43.63\pm  0.07$&$ 7.65_{- 0.37}^{+ 0.40}$&$ 6.97_{- 0.37}^{+ 0.40}$&$ 7.66_{- 0.44}^{+ 0.46}$&$ 6.90_{- 0.40}^{+ 0.42}$&2,3,4      \\
Mrk79&                   &  &              $ 16.1_{-  6.6}^{+  6.6}$&$ 43.74\pm  0.07$&$ 7.85_{- 0.18}^{+ 0.18}$&$ 7.21_{- 0.18}^{+ 0.18}$&$ 7.75_{- 0.18}^{+ 0.18}$&$ 6.95_{- 0.18}^{+ 0.18}$&2,3,4      \\
Mrk79&                   &  &              $ 16.0_{-  5.8}^{+  6.4}$&$ 43.66\pm  0.07$&$ 7.85_{- 0.16}^{+ 0.17}$&$ 7.22_{- 0.16}^{+ 0.17}$&$ 7.93_{- 0.18}^{+ 0.19}$&$ 7.03_{- 0.16}^{+ 0.18}$&2,3,4      \\
\bfseries Mrk79&         CB&$130  \pm12  $&$ 15.2_{-  5.0}^{+  5.2}$&$ 43.68\pm  0.07$&$ 7.83_{- 0.13}^{+ 0.14}$&$ 7.20_{- 0.14}^{+ 0.15}$&$ 7.83_{- 0.18}^{+ 0.18}$&$ 6.99_{- 0.13}^{+ 0.13}$&1          \\
Mrk110&                  &  &              $ 24.3_{-  8.3}^{+  5.5}$&$ 43.68\pm  0.04$&$ 7.05_{- 0.15}^{+ 0.10}$&$ 6.64_{- 0.15}^{+ 0.10}$&$ 7.02_{- 0.48}^{+ 0.47}$&$ 6.83_{- 0.18}^{+ 0.14}$&2,3,4      \\
Mrk110&                  &  &              $ 20.4_{-  6.3}^{+ 10.5}$&$ 43.75\pm  0.04$&$ 7.04_{- 0.13}^{+ 0.22}$&$ 6.56_{- 0.13}^{+ 0.22}$&$ 6.88_{- 0.35}^{+ 0.40}$&$ 6.69_{- 0.15}^{+ 0.24}$&2,3,4      \\
Mrk110&                  &  &              $ 33.3_{- 10.0}^{+ 14.9}$&$ 43.53\pm  0.05$&$ 7.22_{- 0.13}^{+ 0.19}$&$ 6.80_{- 0.13}^{+ 0.19}$&$ 7.18_{- 0.13}^{+ 0.20}$&$ 6.57_{- 0.13}^{+ 0.20}$&2,3,4      \\
\bfseries Mrk110&        E& $91   \pm7   $&$ 25.5_{-  7.7}^{+  7.7}$&$ 43.67\pm  0.11$&$ 7.10_{- 0.13}^{+ 0.13}$&$ 6.67_{- 0.14}^{+ 0.14}$&$ 7.12_{- 0.18}^{+ 0.21}$&$ 6.70_{- 0.17}^{+ 0.18}$&1          \\
NGC3227&                 PB&$92   \pm6   $&$  3.8_{-  0.8}^{+  0.8}$&$ 42.24\pm  0.11$&$ 7.09_{- 0.10}^{+ 0.10}$&$ 6.32_{- 0.10}^{+ 0.10}$&$ 6.97_{- 0.10}^{+ 0.09}$&$ 6.14_{- 0.10}^{+ 0.09}$&1,4,6,7    \\
NGC3516&                 PB&$181  \pm5   $&$ 11.7_{-  1.5}^{+  1.0}$&$ 42.79\pm  0.20$&$ 7.82_{- 0.07}^{+ 0.06}$&$ 7.04_{- 0.07}^{+ 0.06}$&$ 7.79_{- 0.06}^{+ 0.04}$&$ 6.76_{- 0.06}^{+ 0.04}$&1,4,6,7    \\
SBS1116+583A&            PB&$92   \pm4   $&$  2.3_{-  0.5}^{+  0.6}$&$ 42.14\pm  0.23$&$ 6.78_{- 0.10}^{+ 0.12}$&$ 6.04_{- 0.09}^{+ 0.12}$&$ 6.77_{- 0.28}^{+ 0.29}$&$ 6.02_{- 0.14}^{+ 0.16}$&1,3,4,8    \\
Arp151&                  CB&$118  \pm4   $&$  4.0_{-  0.7}^{+  0.5}$&$ 42.55\pm  0.10$&$ 6.87_{- 0.08}^{+ 0.06}$&$ 6.50_{- 0.07}^{+ 0.05}$&$ 6.64_{- 0.09}^{+ 0.07}$&$ 6.09_{- 0.08}^{+ 0.06}$&1,3,4,8    \\
NGC3783&                 PB&$95   \pm10  $&$ 10.2_{-  2.3}^{+  3.3}$&$ 42.56\pm  0.18$&$ 7.45_{- 0.10}^{+ 0.14}$&$ 6.75_{- 0.10}^{+ 0.14}$&$ 7.28_{- 0.18}^{+ 0.20}$&$ 6.79_{- 0.12}^{+ 0.16}$&1,2,3,4    \\
Mrk1310&                 CB&$84   \pm5   $&$  3.7_{-  0.6}^{+  0.6}$&$ 42.29\pm  0.14$&$ 6.62_{- 0.07}^{+ 0.07}$&$ 6.02_{- 0.08}^{+ 0.08}$&$ 6.26_{- 0.15}^{+ 0.15}$&$ 5.61_{- 0.17}^{+ 0.17}$&1,3,9      \\
NGC4051&                 &  &              $  5.8_{-  2.2}^{+  2.2}$&$ 41.28\pm  0.15$&$ 6.38_{- 0.16}^{+ 0.16}$&$ 6.41_{- 0.16}^{+ 0.16}$&$ 6.11_{- 0.19}^{+ 0.19}$&$ 5.52_{- 0.18}^{+ 0.18}$&2,10,11    \\
NGC4051&                 &  &              $  1.9_{-  0.5}^{+  0.5}$&$ 41.96\pm  0.19$&$ 5.37_{- 0.11}^{+ 0.12}$&$ 5.60_{- 0.12}^{+ 0.12}$&$ 5.50_{- 0.13}^{+ 0.14}$&$ 5.59_{- 0.12}^{+ 0.13}$&3,4,12     \\
NGC4051&                 &  &              $  2.9_{-  1.3}^{+  0.9}$&$ 41.85\pm  0.15$&$ 5.52_{- 0.20}^{+ 0.13}$&$ 5.09_{- 0.20}^{+ 0.13}$&$ 5.70_{- 0.20}^{+ 0.13}$&$ 5.13_{- 0.21}^{+ 0.14}$&4,13       \\
\bfseries NGC4051&       PB&$89   \pm3   $&$  2.8_{-  1.7}^{+  1.7}$&$ 41.66\pm  0.38$&$ 5.67_{- 0.53}^{+ 0.53}$&$ 5.68_{- 0.59}^{+ 0.59}$&$ 5.70_{- 0.31}^{+ 0.31}$&$ 5.46_{- 0.25}^{+ 0.25}$&1          \\
NGC4151&                 CB&$97   \pm3   $&$  6.6_{-  0.8}^{+  1.1}$&$ 42.09\pm  0.21$&$ 7.72_{- 0.05}^{+ 0.08}$&$ 6.84_{- 0.05}^{+ 0.07}$&$ 7.36_{- 0.12}^{+ 0.13}$&$ 6.67_{- 0.05}^{+ 0.08}$&1,2,3,4    \\
Mrk202&                  PB&$78   \pm3   $&$  3.0_{-  1.1}^{+  1.7}$&$ 42.26\pm  0.14$&$ 6.11_{- 0.16}^{+ 0.24}$&$ 5.65_{- 0.16}^{+ 0.25}$&$ 6.04_{- 0.22}^{+ 0.29}$&$ 5.41_{- 0.18}^{+ 0.26}$&1,3,4,8    \\
NGC4253&                 PB&$93   \pm32  $&$  6.2_{-  1.2}^{+  1.6}$&$ 42.57\pm  0.12$&$ 6.49_{- 0.09}^{+ 0.12}$&$ 6.15_{- 0.09}^{+ 0.12}$&$ 5.92_{- 1.30}^{+ 1.30}$&$ 5.50_{- 0.37}^{+ 0.38}$&1,3,4,8    \\
PG1229+204&              PB&$162  \pm32  $&$ 37.8_{- 15.3}^{+ 27.6}$&$ 43.70\pm  0.05$&$ 8.03_{- 0.17}^{+ 0.31}$&$ 7.28_{- 0.17}^{+ 0.31}$&$ 7.93_{- 0.19}^{+ 0.32}$&$ 7.15_{- 0.19}^{+ 0.32}$&1,2,3,4    \\
NGC4593&                 PB&$135  \pm6   $&$  3.7_{-  0.8}^{+  0.8}$&$ 42.87\pm  0.18$&$ 7.28_{- 0.09}^{+ 0.09}$&$ 6.37_{- 0.09}^{+ 0.09}$&$ 7.10_{- 0.12}^{+ 0.12}$&$ 6.25_{- 0.09}^{+ 0.09}$&1,2,3,14   \\
NGC4748&                 PB&$105  \pm13  $&$  5.6_{-  2.2}^{+  1.6}$&$ 42.56\pm  0.12$&$ 6.61_{- 0.17}^{+ 0.13}$&$ 6.04_{- 0.17}^{+ 0.13}$&$ 6.20_{- 0.21}^{+ 0.18}$&$ 5.67_{- 0.21}^{+ 0.17}$&1,3,4,8    \\
NGC5273&                 E& $74.1 \pm3.7 $&$  2.2_{-  1.6}^{+  1.2}$&$ 41.54\pm  0.16$&$ 7.14_{- 0.31}^{+ 0.23}$&$ 6.16_{- 0.31}^{+ 0.23}$&$ 6.96_{- 0.32}^{+ 0.24}$&$ 6.01_{- 0.32}^{+ 0.24}$&4,15       \\
Mrk279&                  PB&$197  \pm12  $&$ 16.7_{-  3.9}^{+  3.9}$&$ 43.71\pm  0.07$&$ 7.97_{- 0.10}^{+ 0.10}$&$ 7.03_{- 0.10}^{+ 0.10}$&$ 7.57_{- 0.13}^{+ 0.13}$&$ 6.82_{- 0.12}^{+ 0.12}$&1,2,3,4    \\
PG1411+442&              CB&$209  \pm30  $&$124.3_{- 61.7}^{+ 61.0}$&$ 44.56\pm  0.02$&$ 8.28_{- 0.21}^{+ 0.21}$&$ 7.88_{- 0.21}^{+ 0.21}$&$ 8.14_{- 0.25}^{+ 0.25}$&$ 7.80_{- 0.23}^{+ 0.23}$&1,2,3,4    \\
NGC5548&                 &  &              $ 19.7_{-  1.5}^{+  1.5}$&$ 43.39\pm  0.10$&$ 7.92_{- 0.03}^{+ 0.03}$&$ 7.16_{- 0.03}^{+ 0.03}$&$ 7.80_{- 0.05}^{+ 0.05}$&$ 7.04_{- 0.04}^{+ 0.04}$&15         \\
NGC5548&                 &  &              $ 18.6_{-  2.3}^{+  2.1}$&$ 43.14\pm  0.11$&$ 8.03_{- 0.06}^{+ 0.05}$&$ 7.25_{- 0.05}^{+ 0.05}$&$ 7.90_{- 0.08}^{+ 0.08}$&$ 7.11_{- 0.07}^{+ 0.06}$&15         \\
NGC5548&                 &  &              $ 15.9_{-  2.5}^{+  2.9}$&$ 43.35\pm  0.09$&$ 7.93_{- 0.07}^{+ 0.08}$&$ 7.18_{- 0.07}^{+ 0.08}$&$ 8.01_{- 0.08}^{+ 0.09}$&$ 7.13_{- 0.08}^{+ 0.09}$&15         \\
NGC5548&                 &  &              $ 11.0_{-  2.0}^{+  1.9}$&$ 43.07\pm  0.11$&$ 7.89_{- 0.08}^{+ 0.08}$&$ 7.32_{- 0.08}^{+ 0.08}$&$ 7.84_{- 0.08}^{+ 0.08}$&$ 7.04_{- 0.09}^{+ 0.08}$&15         \\
NGC5548&                 &  &              $ 13.0_{-  1.4}^{+  1.6}$&$ 43.32\pm  0.10$&$ 7.95_{- 0.05}^{+ 0.05}$&$ 7.19_{- 0.05}^{+ 0.05}$&$\cdots$&                 $\cdots$&                 15         \\
NGC5548&                 &  &              $ 13.4_{-  4.3}^{+  3.8}$&$ 43.38\pm  0.09$&$ 8.15_{- 0.14}^{+ 0.12}$&$ 7.34_{- 0.14}^{+ 0.12}$&$ 8.13_{- 0.15}^{+ 0.13}$&$ 7.34_{- 0.14}^{+ 0.13}$&15         \\
NGC5548&                 &  &              $ 21.7_{-  2.6}^{+  2.6}$&$ 43.52\pm  0.09$&$ 8.31_{- 0.05}^{+ 0.05}$&$ 7.40_{- 0.05}^{+ 0.05}$&$ 8.20_{- 0.07}^{+ 0.07}$&$ 7.33_{- 0.07}^{+ 0.07}$&15         \\
NGC5548&                 &  &              $ 16.4_{-  1.1}^{+  1.2}$&$ 43.43\pm  0.09$&$ 8.15_{- 0.03}^{+ 0.03}$&$ 7.22_{- 0.03}^{+ 0.03}$&$ 8.02_{- 0.06}^{+ 0.06}$&$ 7.12_{- 0.04}^{+ 0.04}$&15         \\
NGC5548&                 &  &              $ 17.5_{-  1.6}^{+  2.0}$&$ 43.24\pm  0.10$&$ 8.13_{- 0.04}^{+ 0.05}$&$ 7.21_{- 0.04}^{+ 0.05}$&$ 8.02_{- 0.07}^{+ 0.07}$&$ 7.10_{- 0.05}^{+ 0.06}$&15         \\
NGC5548&                 &  &              $ 26.5_{-  2.2}^{+  4.3}$&$ 43.59\pm  0.09$&$ 8.29_{- 0.04}^{+ 0.07}$&$ 7.33_{- 0.04}^{+ 0.07}$&$ 8.04_{- 0.10}^{+ 0.12}$&$ 7.19_{- 0.05}^{+ 0.08}$&15         \\
NGC5548&                 &  &              $ 24.8_{-  3.0}^{+  3.2}$&$ 43.51\pm  0.09$&$ 8.28_{- 0.05}^{+ 0.06}$&$ 7.30_{- 0.05}^{+ 0.06}$&$ 8.29_{- 0.06}^{+ 0.06}$&$ 7.28_{- 0.05}^{+ 0.06}$&15         \\
NGC5548&                 &  &              $  6.5_{-  3.7}^{+  5.7}$&$ 43.11\pm  0.11$&$ 7.69_{- 0.24}^{+ 0.38}$&$ 6.71_{- 0.25}^{+ 0.38}$&$ 7.65_{- 0.25}^{+ 0.38}$&$ 6.69_{- 0.25}^{+ 0.38}$&15         \\
NGC5548&                 &  &              $ 14.3_{-  7.3}^{+  5.9}$&$ 43.11\pm  0.11$&$ 8.07_{- 0.22}^{+ 0.18}$&$ 7.43_{- 0.22}^{+ 0.18}$&$ 8.04_{- 0.22}^{+ 0.18}$&$ 7.12_{- 0.22}^{+ 0.18}$&15         \\
NGC5548&                 &  &              $  6.3_{-  2.3}^{+  2.6}$&$ 42.96\pm  0.13$&$ 7.70_{- 0.16}^{+ 0.18}$&$ 6.94_{- 0.23}^{+ 0.25}$&$\cdots$&                 $\cdots$&                 15         \\
NGC5548&                 &  &              $  4.2_{-  1.3}^{+  0.9}$&$ 43.01\pm  0.11$&$ 8.12_{- 0.13}^{+ 0.09}$&$ 7.17_{- 0.13}^{+ 0.09}$&$ 8.01_{- 0.22}^{+ 0.20}$&$ 7.17_{- 0.15}^{+ 0.11}$&15         \\
NGC5548&                 &  &              $ 12.4_{-  3.8}^{+  2.7}$&$ 42.99\pm  0.11$&$ 8.50_{- 0.14}^{+ 0.10}$&$ 7.63_{- 0.14}^{+ 0.10}$&$ 7.75_{- 0.13}^{+ 0.10}$&$ 6.90_{- 0.13}^{+ 0.10}$&15,16      \\
NGC5548&                 &  &              $  7.2_{-  0.3}^{+  1.3}$&$ 43.21\pm  0.09$&$ 8.14_{- 0.04}^{+ 0.08}$&$ 7.20_{- 0.07}^{+ 0.10}$&$ 8.10_{- 0.03}^{+ 0.08}$&$ 7.14_{- 0.09}^{+ 0.11}$&15         \\
NGC5548&                 &  &              $  4.2_{-  0.4}^{+  0.4}$&$ 43.45\pm  0.09$&$ 7.87_{- 0.06}^{+ 0.06}$&$ 7.05_{- 0.06}^{+ 0.06}$&$ 7.93_{- 0.06}^{+ 0.06}$&$ 7.18_{- 0.14}^{+ 0.14}$&17         \\
\bfseries NGC5548&       CB&$195  \pm13  $&$ 14.2_{-  7.9}^{+  7.9}$&$ 43.30\pm  0.19$&$ 8.08_{- 0.16}^{+ 0.16}$&$ 7.23_{- 0.10}^{+ 0.10}$&$ 8.01_{- 0.16}^{+ 0.17}$&$ 7.12_{- 0.11}^{+ 0.11}$&1          \\
PG1426+015&              E& $217  \pm15  $&$ 95.0_{- 37.1}^{+ 29.9}$&$ 44.63\pm  0.02$&$ 8.97_{- 0.17}^{+ 0.14}$&$ 8.19_{- 0.17}^{+ 0.14}$&$ 8.87_{- 0.24}^{+ 0.22}$&$ 8.34_{- 0.18}^{+ 0.16}$&1,2,3,4    \\
\hline
\end{tabular}
\end{lrbox}
\scalebox{0.85}{\usebox{\tablebox}}
\end{table*}

\begin{table*}
\centering
\label{tab2}
\begin{lrbox}{\tablebox}
\begin{tabular}{llllllllllllll}
\hline
Name&Bulge&\sst&$\tau$&$\log \lv$&$\log \vpfm$&$\log \vpsm$&$\log \vpfr$&$\log \vpsr$&Ref.\\
         &Type&(\kms)&(days)&$\log \ergs$&($\log \msun$)&($\log \msun$)&($\log \msun$)&($\log \msun$)&\\
 (1)&(2)&(3)&(4)&(5)&(6)&(7)&(8)&(9)&(10)\\
\hline
Mrk817&                  &  &              $ 19.0_{-  3.7}^{+  3.9}$&$ 43.79\pm  0.05$&$ 7.92_{- 0.08}^{+ 0.09}$&$ 7.16_{- 0.08}^{+ 0.09}$&$ 7.66_{- 0.13}^{+ 0.13}$&$ 6.86_{- 0.10}^{+ 0.10}$&2,3,4      \\
Mrk817&                  &  &              $ 15.3_{-  3.5}^{+  3.7}$&$ 43.67\pm  0.05$&$ 7.91_{- 0.10}^{+ 0.10}$&$ 7.12_{- 0.10}^{+ 0.10}$&$ 7.86_{- 0.14}^{+ 0.14}$&$ 7.06_{- 0.11}^{+ 0.11}$&2,3,4      \\
Mrk817&                  &  &              $ 33.6_{-  7.6}^{+  6.5}$&$ 43.67\pm  0.05$&$ 8.17_{- 0.10}^{+ 0.08}$&$ 7.50_{- 0.10}^{+ 0.08}$&$ 7.96_{- 0.25}^{+ 0.24}$&$ 7.29_{- 0.13}^{+ 0.11}$&2,3,4      \\
Mrk817&                  &  &              $ 14.0_{-  3.5}^{+  3.4}$&$ 43.84\pm  0.05$&$ 7.94_{- 0.11}^{+ 0.10}$&$ 7.05_{- 0.11}^{+ 0.10}$&$\cdots$&                 $\cdots$&                 6,19       \\
\bfseries Mrk817&        PB&$120  \pm15  $&$ 19.9_{-  8.2}^{+  8.2}$&$ 43.74\pm  0.09$&$ 7.99_{- 0.14}^{+ 0.14}$&$ 7.22_{- 0.21}^{+ 0.21}$&$ 7.78_{- 0.17}^{+ 0.17}$&$ 7.04_{- 0.22}^{+ 0.22}$&1          \\
Mrk290&                  CB&$110  \pm5   $&$  8.7_{-  1.0}^{+  1.2}$&$ 43.17\pm  0.06$&$ 7.55_{- 0.07}^{+ 0.07}$&$ 6.73_{- 0.07}^{+ 0.07}$&$ 7.49_{- 0.06}^{+ 0.07}$&$ 6.64_{- 0.06}^{+ 0.06}$&1,2,6,7,19 \\
PG1617+175&              E& $201  \pm37  $&$ 71.5_{- 33.7}^{+ 29.6}$&$ 44.39\pm  0.02$&$ 8.79_{- 0.20}^{+ 0.18}$&$ 7.87_{- 0.20}^{+ 0.18}$&$ 8.49_{- 0.27}^{+ 0.25}$&$ 7.98_{- 0.21}^{+ 0.19}$&1,2,3,4    \\
3C390.3&                 &  &              $ 23.6_{-  6.7}^{+  6.2}$&$ 43.68\pm  0.10$&$ 8.87_{- 0.12}^{+ 0.11}$&$ 7.81_{- 0.12}^{+ 0.11}$&$ 8.66_{- 0.15}^{+ 0.14}$&$ 7.65_{- 0.12}^{+ 0.12}$&2,3,4      \\
3C390.3&                 &  &              $ 46.4_{-  3.2}^{+  3.6}$&$ 44.50\pm  0.03$&$ 9.20_{- 0.03}^{+ 0.03}$&$ 8.42_{- 0.03}^{+ 0.03}$&$ 9.03_{- 0.14}^{+ 0.14}$&$ 8.43_{- 0.05}^{+ 0.06}$&4,5        \\
\bfseries 3C390.3&       E& $273  \pm16  $&$ 44.3_{- 11.1}^{+ 11.2}$&$ 44.43\pm  0.32$&$ 9.18_{- 0.12}^{+ 0.12}$&$ 8.38_{- 0.22}^{+ 0.22}$&$ 8.86_{- 0.28}^{+ 0.28}$&$ 8.30_{- 0.42}^{+ 0.42}$&1          \\
NGC6814&                 PB&$95   \pm3   $&$  6.6_{-  0.9}^{+  0.9}$&$ 42.12\pm  0.28$&$ 7.16_{- 0.06}^{+ 0.06}$&$ 6.68_{- 0.06}^{+ 0.06}$&$ 7.14_{- 0.10}^{+ 0.10}$&$ 6.53_{- 0.08}^{+ 0.08}$&1,3,4,8    \\
Mrk509&                  E& $184  \pm12  $&$ 79.6_{-  5.4}^{+  6.1}$&$ 44.19\pm  0.05$&$ 8.15_{- 0.03}^{+ 0.03}$&$ 7.57_{- 0.03}^{+ 0.03}$&$ 8.06_{- 0.04}^{+ 0.05}$&$ 7.40_{- 0.03}^{+ 0.04}$&1,2,3,4    \\
PG2130+099&              PB&$163  \pm19  $&$  9.6_{-  1.2}^{+  1.2}$&$ 44.20\pm  0.03$&$ 7.05_{- 0.09}^{+ 0.09}$&$ 6.76_{- 0.05}^{+ 0.05}$&$ 6.92_{- 0.07}^{+ 0.07}$&$ 6.79_{- 0.06}^{+ 0.06}$&1,4,6,7,19 \\
NGC7469&                 &  &              $ 10.8_{-  1.3}^{+  3.4}$&$ 43.51\pm  0.11$&$ 7.60_{- 0.05}^{+ 0.14}$&$ 6.40_{- 0.05}^{+ 0.14}$&$ 6.38_{- 0.09}^{+ 0.15}$&$ 6.53_{- 0.10}^{+ 0.16}$&7,19       \\
NGC7469&                 &  &              $  4.5_{-  0.8}^{+  0.7}$&$ 43.32\pm  0.12$&$ 6.42_{- 0.08}^{+ 0.07}$&$ 6.41_{- 0.08}^{+ 0.07}$&$ 6.62_{- 0.20}^{+ 0.19}$&$ 6.27_{- 0.14}^{+ 0.14}$&2,3        \\
\bfseries NGC7469&       PB&$131  \pm5   $&$  6.2_{-  3.8}^{+  3.9}$&$ 43.42\pm  0.16$&$ 6.87_{- 0.82}^{+ 0.82}$&$ 6.41_{- 0.05}^{+ 0.07}$&$ 6.44_{- 0.17}^{+ 0.19}$&$ 6.41_{- 0.21}^{+ 0.22}$&1          \\
Mrk50&                   CB&$109  \pm14  $&$  8.7_{-  1.5}^{+  1.6}$&$ 42.83\pm  0.06$&$ 7.46_{- 0.08}^{+ 0.08}$&$ 6.84_{- 0.08}^{+ 0.08}$&$ 7.28_{- 0.08}^{+ 0.09}$&$ 6.84_{- 0.09}^{+ 0.09}$&1,21       \\
MCG6-30-15&              CB&$109  \pm9   $&$  5.3_{-  1.8}^{+  1.9}$&$ 41.64\pm  0.11$&$ 6.60_{- 0.15}^{+ 0.16}$&$ 5.99_{- 0.15}^{+ 0.15}$&$ 6.32_{- 0.29}^{+ 0.30}$&$ 5.66_{- 0.18}^{+ 0.19}$&23         \\
UGC06728&                CB&$52   \pm5   $&$  1.4_{-  0.8}^{+  0.7}$&$ 41.86\pm  0.08$&$ 5.55_{- 0.25}^{+ 0.22}$&$ 5.20_{- 0.25}^{+ 0.22}$&$ 5.67_{- 0.27}^{+ 0.25}$&$ 5.22_{- 0.27}^{+ 0.24}$&24         \\
MCG+06-26-012&           PB&$112  \pm15  $&$ 24.0_{-  4.8}^{+  8.4}$&$ 42.67\pm  0.11$&$ 6.92_{- 0.10}^{+ 0.16}$&$ 6.46_{- 0.09}^{+ 0.15}$&$\cdots$&                 $\cdots$&                 7,19,24,25 \\
RMID17&                  &  $191.4\pm3.7 $&$ 25.5_{-  5.8}^{+ 10.9}$&$ 43.94\pm  0.00$&$ 9.12_{- 0.10}^{+ 0.18}$&$ 8.38_{- 0.10}^{+ 0.18}$&$ 8.48_{- 0.10}^{+ 0.18}$&$ 8.27_{- 0.10}^{+ 0.18}$&26,27      \\
RMID33&                  &  $182.4\pm21.7$&$ 26.5_{-  8.8}^{+  9.9}$&$ 44.14\pm  0.00$&$ 6.77_{- 0.14}^{+ 0.16}$&$ 6.49_{- 0.14}^{+ 0.16}$&$ 7.14_{- 0.19}^{+ 0.21}$&$ 6.58_{- 0.15}^{+ 0.16}$&26,27      \\
RMID177&                 &  $171.5\pm10.7$&$ 10.1_{-  2.7}^{+ 12.5}$&$ 43.99\pm  0.00$&$ 7.74_{- 0.12}^{+ 0.53}$&$ 7.10_{- 0.11}^{+ 0.53}$&$ 7.68_{- 0.12}^{+ 0.53}$&$ 6.91_{- 0.12}^{+ 0.53}$&26,27      \\
RMID191&                 &  $152  \pm8.5 $&$  8.5_{-  1.4}^{+  2.5}$&$ 43.68\pm  0.01$&$ 6.46_{- 0.09}^{+ 0.14}$&$ 6.07_{- 0.07}^{+ 0.13}$&$ 6.81_{- 0.08}^{+ 0.13}$&$ 6.25_{- 0.07}^{+ 0.13}$&26,27      \\
RMID229&                 &  $130.2\pm8.7 $&$ 16.2_{-  4.5}^{+  2.9}$&$ 43.58\pm  0.00$&$ 7.47_{- 0.13}^{+ 0.09}$&$ 6.97_{- 0.12}^{+ 0.08}$&$ 7.25_{- 0.16}^{+ 0.13}$&$ 7.00_{- 0.12}^{+ 0.08}$&26,27      \\
RMID267&                 &  $97.1 \pm9   $&$ 20.4_{-  2.0}^{+  2.5}$&$ 44.12\pm  0.00$&$ 7.45_{- 0.04}^{+ 0.05}$&$ 6.83_{- 0.04}^{+ 0.05}$&$ 7.20_{- 0.05}^{+ 0.06}$&$ 6.76_{- 0.05}^{+ 0.06}$&26,27      \\
RMID300&                 &  $109.4\pm11.9$&$ 30.4_{-  8.3}^{+  3.9}$&$ 44.52\pm  0.02$&$ 7.42_{- 0.12}^{+ 0.06}$&$ 6.90_{- 0.12}^{+ 0.06}$&$ 7.59_{- 0.13}^{+ 0.07}$&$ 6.95_{- 0.12}^{+ 0.06}$&26,27      \\
RMID301&                 &  $176.9\pm10.1$&$ 12.8_{-  4.5}^{+  5.7}$&$ 44.09\pm  0.00$&$ 8.95_{- 0.15}^{+ 0.19}$&$ 8.09_{- 0.15}^{+ 0.19}$&$ 8.44_{- 0.15}^{+ 0.19}$&$ 7.99_{- 0.15}^{+ 0.19}$&26,27      \\
RMID305&                 &  $150.5\pm7.7 $&$ 53.5_{-  4.0}^{+  4.2}$&$ 44.22\pm  0.00$&$ 7.85_{- 0.03}^{+ 0.03}$&$ 7.75_{- 0.03}^{+ 0.03}$&$ 8.02_{- 0.04}^{+ 0.04}$&$ 7.67_{- 0.04}^{+ 0.04}$&26,27      \\
RMID320&                 &  $66.4 \pm4.6 $&$ 25.2_{-  5.7}^{+  4.7}$&$ 43.46\pm  0.00$&$ 7.88_{- 0.10}^{+ 0.08}$&$ 7.08_{- 0.10}^{+ 0.08}$&$ 7.56_{- 0.10}^{+ 0.08}$&$ 7.02_{- 0.10}^{+ 0.08}$&26,27      \\
RMID338&                 &  $83.3 \pm8.3 $&$ 10.7_{-  4.4}^{+  5.6}$&$ 43.40\pm  0.00$&$ 7.66_{- 0.21}^{+ 0.25}$&$ 7.17_{- 0.18}^{+ 0.23}$&$ 7.74_{- 0.18}^{+ 0.23}$&$ 7.04_{- 0.18}^{+ 0.23}$&26,27      \\
RMID377&                 &  $115.3\pm4.6 $&$  5.9_{-  0.6}^{+  0.4}$&$ 43.40\pm  0.00$&$ 7.16_{- 0.04}^{+ 0.03}$&$ 6.49_{- 0.04}^{+ 0.03}$&$ 7.57_{- 0.06}^{+ 0.05}$&$ 6.57_{- 0.05}^{+ 0.03}$&26,27      \\
RMID392&                 &  $77.2 \pm25.6$&$ 14.2_{-  3.0}^{+  3.7}$&$ 44.31\pm  0.01$&$ 7.54_{- 0.10}^{+ 0.12}$&$ 7.43_{- 0.09}^{+ 0.11}$&$ 8.51_{- 0.09}^{+ 0.11}$&$ 7.57_{- 0.09}^{+ 0.11}$&26,27      \\
RMID399&                 &  $187.2\pm7.8 $&$ 35.8_{- 10.3}^{+  1.1}$&$ 44.02\pm  0.00$&$ 7.70_{- 0.13}^{+ 0.02}$&$ 7.15_{- 0.12}^{+ 0.02}$&$ 7.67_{- 0.13}^{+ 0.04}$&$ 7.26_{- 0.13}^{+ 0.02}$&26,27      \\
RMID457&                 &  $110  \pm18.4$&$ 15.6_{-  5.1}^{+  3.2}$&$ 43.40\pm  0.01$&$ 8.10_{- 0.15}^{+ 0.10}$&$ 7.43_{- 0.14}^{+ 0.09}$&$ 8.23_{- 0.14}^{+ 0.09}$&$ 7.37_{- 0.14}^{+ 0.09}$&26,27      \\
RMID601&                 &  $214.9\pm19.2$&$ 11.6_{-  4.6}^{+  8.6}$&$ 44.15\pm  0.00$&$ 8.77_{- 0.17}^{+ 0.32}$&$ 8.01_{- 0.17}^{+ 0.32}$&$ 8.56_{- 0.17}^{+ 0.32}$&$ 7.80_{- 0.17}^{+ 0.32}$&26,27      \\
RMID622&                 &  $122.9\pm9.2 $&$ 49.1_{-  2.0}^{+ 11.1}$&$ 44.31\pm  0.00$&$ 7.80_{- 0.02}^{+ 0.10}$&$ 7.25_{- 0.02}^{+ 0.10}$&$ 8.00_{- 0.05}^{+ 0.11}$&$ 7.29_{- 0.03}^{+ 0.10}$&26,27      \\
RMID634&                 &  $119.4\pm20.9$&$ 17.6_{-  7.4}^{+  8.6}$&$ 44.04\pm  0.00$&$ 6.66_{- 0.18}^{+ 0.21}$&$ 6.59_{- 0.18}^{+ 0.21}$&$ 7.60_{- 0.22}^{+ 0.24}$&$ 6.90_{- 0.18}^{+ 0.21}$&26,27      \\
RMID772&                 &  $136.5\pm3.1 $&$  3.9_{-  0.9}^{+  0.9}$&$ 43.47\pm  0.00$&$ 6.66_{- 0.10}^{+ 0.10}$&$ 5.94_{- 0.10}^{+ 0.10}$&$ 6.52_{- 0.10}^{+ 0.10}$&$ 5.90_{- 0.10}^{+ 0.10}$&26,27      \\
RMID775&                 &  $130.4\pm2.6 $&$ 16.3_{-  6.6}^{+ 13.1}$&$ 43.58\pm  0.00$&$ 7.48_{- 0.17}^{+ 0.35}$&$ 6.90_{- 0.17}^{+ 0.35}$&$ 7.90_{- 0.17}^{+ 0.35}$&$ 7.02_{- 0.17}^{+ 0.35}$&26,27      \\
RMID776&                 &  $112.4\pm1.9 $&$ 10.5_{-  2.2}^{+  1.0}$&$ 43.15\pm  0.00$&$ 7.45_{- 0.09}^{+ 0.04}$&$ 6.66_{- 0.09}^{+ 0.04}$&$ 7.30_{- 0.09}^{+ 0.04}$&$ 6.61_{- 0.09}^{+ 0.04}$&26,27      \\
RMID779&                 &  $57.1 \pm4.9 $&$ 11.8_{-  1.5}^{+  0.7}$&$ 43.12\pm  0.00$&$ 7.21_{- 0.05}^{+ 0.03}$&$ 6.56_{- 0.05}^{+ 0.03}$&$ 7.23_{- 0.06}^{+ 0.03}$&$ 6.52_{- 0.06}^{+ 0.03}$&26,27      \\
RMID781&                 &  $104.7\pm4.3 $&$ 75.2_{-  3.3}^{+  3.2}$&$ 43.64\pm  0.02$&$ 7.97_{- 0.02}^{+ 0.02}$&$ 7.30_{- 0.02}^{+ 0.02}$&$ 8.21_{- 0.03}^{+ 0.03}$&$ 7.24_{- 0.03}^{+ 0.03}$&26,27      \\
RMID782&                 &  $129.5\pm6.7 $&$ 20.0_{-  3.0}^{+  1.1}$&$ 43.94\pm  0.00$&$ 7.57_{- 0.07}^{+ 0.03}$&$ 6.87_{- 0.06}^{+ 0.02}$&$ 7.46_{- 0.08}^{+ 0.05}$&$ 6.85_{- 0.07}^{+ 0.03}$&26,27      \\
RMID790&                 &  $204.4\pm3.1 $&$  5.5_{-  2.1}^{+  5.7}$&$ 43.29\pm  0.00$&$ 8.50_{- 0.16}^{+ 0.45}$&$ 7.70_{- 0.16}^{+ 0.45}$&$ 7.98_{- 0.17}^{+ 0.45}$&$ 7.63_{- 0.16}^{+ 0.45}$&26,27      \\
RMID840&                 &  $164.3\pm3.6 $&$  5.0_{-  1.4}^{+  1.5}$&$ 43.21\pm  0.00$&$ 8.38_{- 0.12}^{+ 0.13}$&$ 7.63_{- 0.12}^{+ 0.13}$&$ 7.63_{- 0.12}^{+ 0.13}$&$ 7.29_{- 0.12}^{+ 0.13}$&26,27      \\

\hline
\end{tabular}
\end{lrbox}
\scalebox{0.85}{\usebox{\tablebox}}
\\
Note. For a object with multiple measurements, we highlighted the name in boldface. \\
Reference: 1: \cite{HK14}, 2: \cite{Co06}, 3: \cite{Be13}, 4: \cite{Du19}, 5: \cite{Gr12}, 6: \cite{Denny2010}, 7: \cite{Du15}, 8: \cite{Be09b}, 9: \cite{Be09a}, 10: \cite{Pe00}, 11: \cite{Be06}, 12: \cite{Denny2009}, 13: \cite{Fa17}, 14: \cite{Denny2006}, 15: \cite{Be14}, 16: \cite{Lu16}, 17: \cite{Du16b}, 18: \cite{Pei17}, 19: \cite{Du16a}, 20: \cite{Pe14}, 21: \cite{Williams2018}, 22: \cite{Be16a}, 23: \cite{Be16b}, 24: \cite{Wang2014}, 25: \cite{Woo15}, 26: \cite{Sh15}, 27: \cite{Gr17}\\
(This table is available in its entirety in machine-readable form.)
\end{table*}

\begin{table*}
\caption{Four kinds of virial factor for Reverberation-mapped AGNs.}
\begin{tabular}{lllllll}
\hline
   VP            & Sample     & $f$                                                         & $\sigma_{\rm int}$(dex) & $\Delta_{\rm CB}$(dex) & $\Delta_{\rm All}$(dex) & Ref. \\
   (1)&(2)&(3)&(4)&(5)&(6)&(7)\\\hline
\vpfm      & 17 E/CBs      & $1.12 \pm 0.17 $                                              & $0.38 \pm 0.08$       & 0.44               & 0.52                & 1   \\
               & \blue{ 62 RM AGNs} & \blue{ $ 1.29  \pm 0.16   $	}				& \blue{ $0.57 \pm 0.05$  }     &                    & \blue{ 0.62} & 1\\
               & \blue{43 RM AGNs$^a$} & \blue{ $ 1.00  \pm 0.16   $	}				& \blue{ $0.61 \pm 0.06$  }     &                     & \blue{ 0.64} & 1\\

                & 14 E/CBs      & $1.3 \pm 0.4 $                                                & $0.44 \pm 0.10$       &                    &                     & 2   \\
                & 16 PBs      & $0.5 \pm 0.2$                                                 & $0.41 \pm 0.09 $      &                    &                     & 2   \\
\multirow{2}{*}{} & \multirow{2}{*}{17 E/CBs }      & \multirow{2}{*}{$-(1.10 \pm 0.40)\log\frac{\fwm}{2000\kms} + (0.50 \pm 0.11)$ } & \multirow{2}{*}{$0.37 \pm 0.06$}        &\multirow{2}{*}{0.39  }               & \multirow{2}{*}{0.39    }             & \multirow{2}{*}{1 }   \\
                \specialrule{0em}{1.5pt}{1.5pt}

\multirow{2}{*}{} & \multirow{2}{*}{\blue{ 62 RM AGNs} }      & \multirow{2}{*}{\blue{ $-(1.05 \pm 0.18)\log\frac{\fwm}{2000\kms} + (0.39 \pm 0.07)$} } & \multirow{2}{*}{\blue{ $0.48 \pm 0.05$}}        &               & \multirow{2}{*}{\blue{0.53}   }             & \multirow{2}{*}{1 }   \\
                \specialrule{0em}{1.5pt}{1.5pt}

\multirow{2}{*}{} & \multirow{2}{*}{\blue{ 43 RM AGNs$^a$} }      & \multirow{2}{*}{\blue{ $-(1.23 \pm 0.17)\log\frac{\fwm}{2000\kms} + (0.53 \pm 0.08)$} } & \multirow{2}{*}{\blue{ $0.36 \pm 0.03$}}        &               & \multirow{2}{*}{\blue{0.60}   }             & \multirow{2}{*}{1 }   \\
                \specialrule{0em}{1.5pt}{1.5pt}

    \multirow{2}{*}{}            & \multirow{2}{*}{34 RM AGNs} & \multirow{2}{*}{ $ -(1.11 \pm 0.27)\log\frac{\fwm}{2000\kms} + (0.48 \pm 0.09) $}  &              \multirow{2}{*}{}            &       \multirow{2}{*}{}                  &  \multirow{2}{*}{0.39  }                &  \multirow{2}{*}{3}       \\
                    \specialrule{0em}{2pt}{2pt}

    \multirow{2}{*}{}            & \multirow{2}{*}{37 AGNs }   &   \multirow{2}{*}{$-(1.17 \pm 0.11)\log\frac{\rm FWHM}{(4550 \pm 1000) \kms}  $  }     &        \multirow{2}{*}{}                &                    &             \multirow{2}{*}{}           &    \multirow{2}{*}{4}   \\
                     &                        &                                                                              &                                &                       &                       &                    \\

\hline
\vpsm & 17 E/CBs     & $5.50 \pm 0.74$                                               & $0.29 \pm 0.01 $      & 0.32               & 0.38                & 1   \\
		& \blue{62 RM AGNs}     & \blue{$5.37 \pm 0.62$}                                               & \blue{$0.51 \pm 0.04 $}      &                & \blue{0.56}                & 1   \\
		& \blue{43 RM AGNs$^a$}     & \blue{$4.17 \pm 0.58$}                                               & \blue{$0.54 \pm 0.05 $}      &                & \blue{0.59}                & 1   \\		
                & 14 E/CBs     & $5.6 \pm 1.3$                                                 &$ 0.29 \pm 0.06  $     &                    &                     & 2   \\
                & 16 PBs     & $1.9 \pm 0.7$                                                 & $0.31 \pm 0.08$       &                    &                     & 2   \\

\hline
\vpfr      & 17 E/CBs     & $1.51 \pm 0.20$                                               & $0.29 \pm 0.02 $      & 0.35               & 0.54                & 1   \\
		& \blue{61 RM AGNs}     & \blue{$1.55 \pm 0.22$}                                               & \blue{$0.59 \pm 0.04 $}      &                & \blue{0.65}                & 1   \\
		& \blue{43 RM AGNs$^a$}     & \blue{$1.10 \pm 0.17$}                                               & \blue{$0.57 \pm 0.06 $}      &                & \blue{0.63}                & 1   \\		
                & 14 E/CBs     & $1.5 \pm 0.4$                                                 & $0.39 \pm 0.07 $      &                    &                     & 2   \\
                & 16 PBs     & $0.7 \pm 0.2$                                                 & $0.37 \pm 0.07$       &                    &                     & 2   \\
                \multirow{2}{*}{} & \multirow{2}{*}{17 E/CBs }      & \multirow{2}{*}{$-(0.50 \pm 0.25)\log\frac{\fwr}{2000\kms} + (0.34 \pm 0.11)$ } & \multirow{2}{*}{$0.46 \pm 0.09$}        &\multirow{2}{*}{0.34  }               & \multirow{2}{*}{0.45    }             & \multirow{2}{*}{1 }   \\
                                \specialrule{0em}{1.5pt}{1.5pt}
                \multirow{2}{*}{} & \multirow{2}{*}{\blue{61 RM AGNs} }      & \multirow{2}{*}{\blue{$-(1.13 \pm 0.16)\log\frac{\fwr}{2000\kms} + (0.45 \pm 0.08)$ }} & \multirow{2}{*}{\blue{$0.51 \pm 0.04$}}        &           & \multirow{2}{*}{\blue{0.56}    }             & \multirow{2}{*}{1 }   \\
                                \specialrule{0em}{1.5pt}{1.5pt}
                \multirow{2}{*}{} & \multirow{2}{*}{\blue{43 RM AGNs$^a$} }      & \multirow{2}{*}{\blue{$-(1.13 \pm 0.24)\log\frac{\fwr}{2000\kms} + (0.45 \pm 0.07)$ }} & \multirow{2}{*}{\blue{$0.51 \pm 0.03$}}        &           & \multirow{2}{*}{\blue{0.61}    }             & \multirow{2}{*}{1 }   \\
                 &                        &                                                                              &                                &                       &                       &                    \\

\hline
\vpsr & 17 E/CBs     & $6.61 \pm 0.81$                                              & $0.29 \pm 0.02$       & 0.34               & 0.47                & 1   \\
                & \blue{61 RM AGNs}      & \blue{$6.46 \pm 0.81$}                                              & \blue{$0.53 \pm 0.04$}       &               & \blue{0.59}            & 1   \\
                & \blue{43 RM AGNs$^a$}      & \blue{$4.68 \pm 0.60$}                                              & \blue{$0.53 \pm 0.06$}       &               & \blue{0.51}            & 1   \\
                & 15 E/CBs     & $6.3 \pm 1.5$                                                 & $0.39 \pm 0.07 $      &                    &                     & 2   \\
                & 16 PBs     & $3.2 \pm 0.7$                                                 & $0.34 \pm 0.06$       &                    &                     & 2   \\
                \hline

\end{tabular}
\\Note. The samples for calibrating $f$ are listed in Column (2). The superscript of a indicates the subsample of 43 RM-AGNs excluding PBs. Column(10) is the reference of the calibration. Reference: 1: This work, 2: \cite{HK14},  3: \cite{Yu19}, 4: \cite{Mejia2018}
\label{tab3}
\end{table*}

\begin{table*}
\caption{
The results of the liner regression fits for the correlations among four kinds of velocity tracer (Fig. \ref{fig5}). $k_1$ and $k_2$ are the intercept and slope that derived from BCES bisector. N is the number of the RM AGNs. 
$\Delta$ is the offset rms in the y-axis. The last two columns are the Spearman correlation coefficient $r_{s}$ and probability of the null hypothesis $p_{\rm null}$.
}
\begin{tabular}{lllllll}
\hline
                                  &N& $k_1$ & $k_2$  & $\Delta$(dex) & $r_s$     & $p_{\rm null}$        \\\hline
$\log\fwm = k_1 +k_2 \log\sr$ & 106&$0.18 \pm 0.25 $  & $1.04 \pm 0.08$  &  0.19         & 0.81 & 0.21E-24 \\
$\log\fwr = k_1 +k_2 \log\sr$ & 106&$0.11 \pm 0.17 $  & $1.06 \pm 0.05$  & 0.13   & 0.89  & 0.21E-35 \\
$\log\fwr = k_1 +k_2 \log\fwm$ & 106 &$-0.10 \pm 0.27 $  & $1.03 \pm 0.08$  & 0.20   & 0.83  & 0.17E-26 \\
$\log\sm =  k_1 +k_2 \log\sr$  & 106 &$0.34 \pm 0.12$   & $0.91 \pm 0.04$  & 0.09              & 0.90  & 0.78E-44\\
$\log\sm =  k_1 +k_2 \log\fwm$&120&$0.19 \pm 0.14$& $0.87 \pm 0.04$& 0.13&0.87&0.31E-37\\
$\log\sm =  k_1 +k_2 \log\fwr$&106&$0.25 \pm 0.16$& $0.86 \pm 0.05$& 0.14&0.83&0.13E-27\\

\hline
\end{tabular}
\label{tab4}
\end{table*}

\begin{table*}
\caption{Statistics of the difference between the \mbh calculated from \sr ($M_{\rm BH,\sigma, rms}$) and the \mbh calculate from other line width parameters. }
\begin{tabular}{lrrrr}
\hline
                                                & \multicolumn{2}{l}{Exclude SDSS-RM} & \multicolumn{2}{l}{ALL Sources} \\\cline{2-5}
                                                & $\Delta$(dex)     & offset(dex)       & $\Delta$(dex)   & offset(dex)     \\\hline
$M_{\rm BH,c}$ vs $M_{\rm BH,\sigma,rms}$           & 0.33            & $0.15 \pm 0.29$     & 0.40          & $0.01 \pm 0.40$  \\
$M_{\rm BH,F,mean}$ vs $M_{\rm BH,\sigma,rms}$      & 0.34            &$ -0.10 \pm 0.33$    & 0.42          & $-0.19 \pm 0.38$  \\
$M_{\rm BH,F,rms}$ vs $M_{\rm BH,\sigma,rms}$       & 0.29            & $-0.09 \pm 0.28$    & 0.27          & $-0.07 \pm 0.26$  \\
$M_{\rm BH,\sigma,mean}$ vs $M_{\rm BH,\sigma,rms}$ & 0.23            & $0.06 \pm 0.22$     & 0.21          & $-0.01 \pm 0.21$ \\\hline
\end{tabular}
\label{tab5}
\end{table*}

\end{document}